\documentclass[12pt,preprint,numberedappendix,twocolappendix]{emulateapj}

\newcommand\bibinc{n}		

\usepackage{subeqnarray}
\usepackage{amsmath}
\usepackage{hyperref}
\usepackage[normalem]{ulem}

\hypersetup{colorlinks=true,linkcolor=black,citecolor=blue,urlcolor=black}

\def\tt{\bf   }

\newcommand{\Eq}[1]{Equation\,(\ref{#1})}

\newcommand{\App}[1]{Appendix~\ref{#1}}
\newcommand{\Sec}[1]{Section~\ref{#1}}

\newcommand{\Fig}[1]{Figure~\ref{#1}}

\begin{document}

\slugcomment{Submitted to The Astrophysical Journal}

\shorttitle{Dayside-Nightside Temperature Differences in Hot Jupiter Atmospheres}
\shortauthors{T.D. Komacek \& A.P. Showman}

\title{Atmospheric Circulation of Hot Jupiters: Dayside-Nightside Temperature Differences}
\author{Thaddeus D. Komacek$^1$ and Adam P. Showman$^1$} \affil{$^1$Department of Planetary Sciences,
 University of Arizona, Tucson, AZ, 85721 \\
\url{tkomacek@lpl.arizona.edu}} 
\begin{abstract}
The full-phase infrared light curves of low-eccentricity hot Jupiters show a trend of increasing dayside-to-nightside brightness temperature difference with increasing equilibrium temperature. Here we present a three-dimensional model that explains this relationship, in order to shed insight on the processes that control heat redistribution in tidally-locked planetary atmospheres. This three-dimensional model combines predictive analytic theory for the atmospheric circulation and dayside-nightside temperature differences over a range of equilibrium temperature, atmospheric composition, and potential frictional drag strengths with numerical solutions of the circulation that verify this analytic theory. This analytic theory shows that the longitudinal propagation of waves mediates dayside-nightside temperature differences in hot Jupiter atmospheres, analogous to the wave adjustment mechanism that regulates the thermal structure in Earth's tropics. These waves can be damped in hot Jupiter atmospheres by either radiative cooling or potential frictional drag. This frictional drag would likely be caused by Lorentz forces in a partially ionized atmosphere threaded by a background magnetic field, and would increase in strength with increasing temperature. Additionally, the amplitude of radiative heating and cooling increases with increasing temperature, and hence both radiative heating/cooling and frictional drag damp waves more efficiently with increasing equilibrium temperature. Radiative heating and cooling play the largest role in controlling dayside-nightside temperature temperature differences in both our analytic theory and numerical simulations, with frictional drag only being important if it is stronger than the Coriolis force. As a result, dayside-nightside temperature differences in hot Jupiter atmospheres increase with increasing stellar irradiation and decrease with increasing pressure. 
\end{abstract}
\section{Introduction}

\begin{figure}
	\centering
	\includegraphics[width=0.5\textwidth]{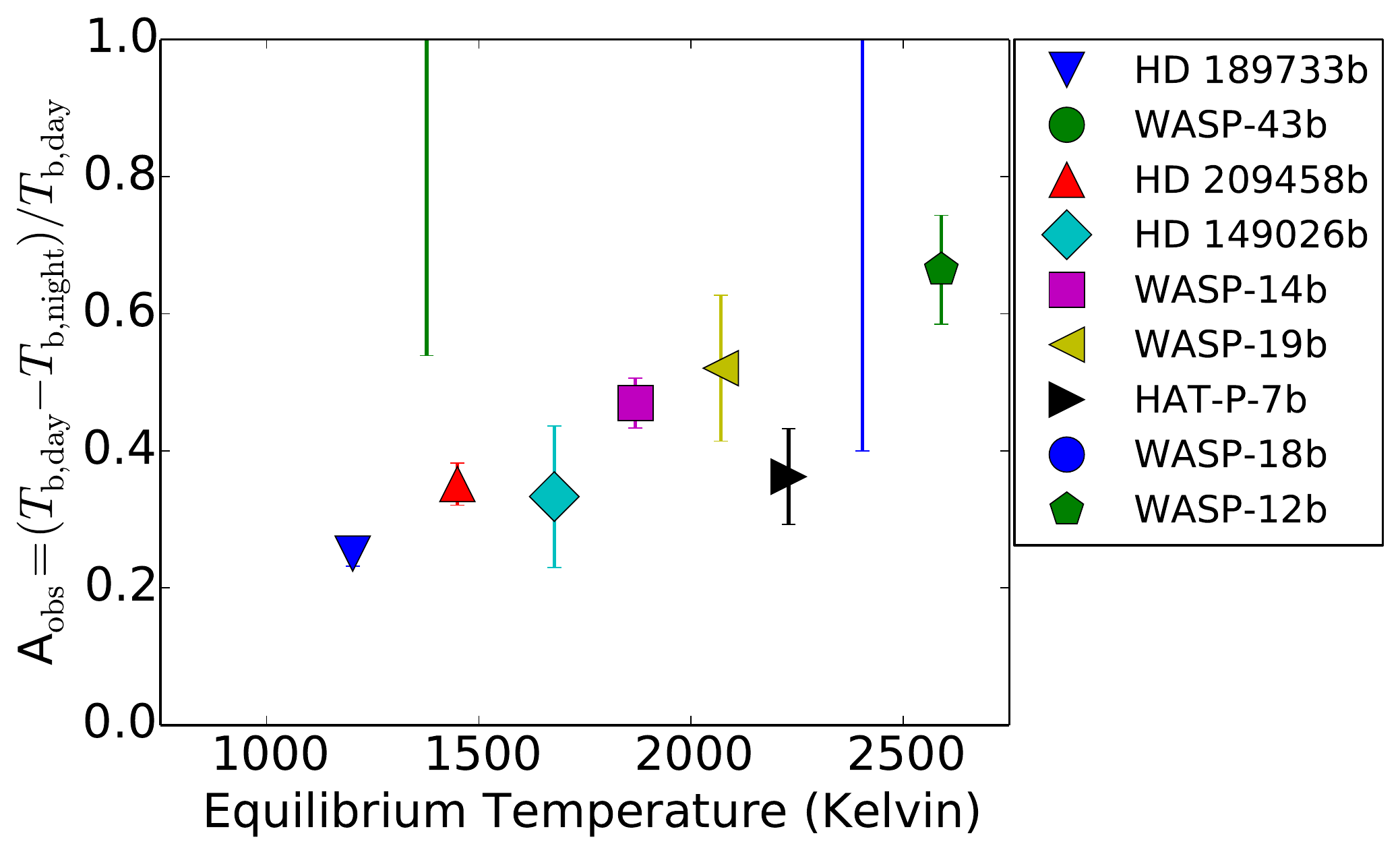}
	\caption{Fractional dayside to nightside brightness
          temperature differences $A_{\mathrm{obs}}$
          vs. global-average equilibrium temperature from
            observations of transiting, low-eccentricity hot
            Jupiters. Here we define the global-average equilibrium
          temperature, $T_{\mathrm{eq}} =
          [F_{\star}/(4\sigma)]^{1/4}$, where $F_{\star}$ is the
          incoming stellar flux to the planet and $\sigma$ is the
          Stefan-Boltzmann constant. Solid points are from the
          full-phase observations of
          \cite{Knutson_2007,Knutson:2009,Knutson:2012} for HD
          189733b, \cite{Crossfield:2012,Zellem:2014} for HD 209458b,
          \cite{Knutson:2009a} for HD 149026b, \cite{Wong:2015} for
          WASP-14b, \cite{Wong:2015a} for WASP-19b and HAT-P-7b, and \cite{Cowan:2012} for WASP-12b. The error bars
          for WASP-43b \citep{Stevenson:2014} and WASP-18b
          \citep{Nymeyer:2011,Maxted:2013} show the lower limit on
          $A_{\mathrm{obs}}$ from the nightside flux upper limits (and
          hence fractional temperature difference lower limits). See
          \App{sec:append1} for the data and method utilized to make
          this figure. There is a clear trend of increasing
          $A_{\mathrm{obs}}$ with increasing equilibrium temperature,
          and hence dayside-nightside temperature differences at the
          photosphere are greater for planets that receive more
          incident flux.}
	\label{fig:Aobs}
\end{figure}

Hot Jupiters, gas giant exoplanets with small semi-major axes and
equilibrium temperatures exceeding $1000\rm\,K$, are the best
characterized class of exoplanets to date. Since the first transit
observations of HD 209458b \citep{Henry:2000,Charbonneau_2000},
infrared (IR) phase curves have been obtained for a variety of objects
(e.g. \citealp{Knutson_2007, Cowan:2007, Borucki:2009, Knutson:2009a,
  Knutson:2009,Cowan:2012,Knutson:2012,Demory_2013,Maxted:2013,
  Stevenson:2014,Zellem:2014,Wong:2015}). Such phase curves allow for
the construction of longitudinally resolved maps of surface
brightness. These maps exhibit a wide diversity, showing that---across
the class of hot Jupiters---the fractional difference between dayside
and nightside flux varies drastically from planet to
planet. 
\noindent \Fig{fig:Aobs} shows the fractional difference
between dayside and nightside brightness temperatures as a function of
equilibrium temperature for the nine low-eccentricity transiting hot
Jupiters with full-phase IR light curve observations. This
fractional difference in dayside-nightside brightness temperature,
$A_{\mathrm{obs}}$, has a value of zero when hot Jupiters are
longitudinally isothermal and unity when the nightside has effectively
no emitted flux relative to the dayside. As seen in \Fig{fig:Aobs},
the fractional dayside-nightside temperature difference increases with
increasing equilibrium temperature. The correlation between fractional
dayside-nightside temperature differences and stellar irradiation
shown in \Fig{fig:Aobs} has also been found by
\cite{Cowan_2011}, \citet{Perez-Becker:2013fv}, and \citet{Schwartz:2015}.

Motivated by these observations, a variety of groups have performed
three-dimensional (3D) numerical simulations of the atmospheric
circulation of hot Jupiters \citep[e.g.][]{showman_2002, Cooper:2005,
  Menou:2009, Showmanetal_2009, Thrastarson:2010, Heng:2011, Heng:2011a,perna_2012, Rauscher_2012, Dobbs-Dixon:2013, Mayne:2014, Showman:2014}.  These
general circulation models (GCMs) generally exhibit day-night
temperature differences ranging from $\sim$200--$1000\rm\,K$
(depending on model details) and fast winds that can exceed several
$\rm km\,s^{-1}$.  When such models include realistic non-grey
radiative transfer, they allow estimates of day-night temperature and
IR flux differences that can be quantitatively compared to phase
curve observations.  Such comparisons are currently the most detailed
for HD 189733b, HD 209458b, and WASP-43b because of the extensive
datasets available for these ``benchmark'' planets
\citep{Showmanetal_2009, Zellem:2014, Kataria:2014}. 

Despite the proliferation of GCM investigations, our understanding of
the underlying {\it dynamical mechanisms} controlling the day-night
temperature differences of hot Jupiters is still in its infancy.  It
is crucial to emphasize that, in and of themselves, GCM simulations do
not automatically imply an understanding: the underlying
dynamics is often sufficiently complex that careful diagnostics and a
hierarchy of simplified models are often necessary
\citep[e.g.,][]{held-2005, Showman_2009}.  The ultimate goal is
not simply matching observations but also understanding physical mechanisms
and constructing a predictive
theory that can quantitatively explain the day-night temperature
differences, horizontal and vertical wind speeds, and other aspects of
the circulation under specified external forcing conditions.  
Taking a step toward such a predictive theory is the primary goal of this paper.

The question of what controls the day-night temperature difference in
hot Jupiter atmospheres has been a subject of intense interest for many years.
Most studies have postulated that the day-night temperature
differences are controlled by a competition between radiation and
atmospheric dynamics---specifically, the tendency of the strong
dayside heating and nightside cooling to create horizontal temperature
differences, and the tendency of the atmospheric circulation to
regulate those temperature differences by transporting thermal energy
from day to night.  Describing this competition using a timescale
comparison, \citet{showman_2002} first suggested that hot Jupiters
would exhibit small fractional dayside-nightside temperature
differences when $\tau_{\mathrm{adv}} \ll \tau_{\mathrm{rad}}$ and
large fractional dayside-nightside temperature differences when
$\tau_{\mathrm{adv}} \gg \tau_{\mathrm{rad}}$. Here,
$\tau_{\mathrm{adv}}$ is the characteristic timescale for the
circulation to advect air parcels horizontally over a hemisphere, and
$\tau_{\mathrm{rad}}$ is the timescale over which radiation modifies
the thermal structure (e.g., the timescale to relax toward the local
radiative equilibrium temperature).  Since then, numerous authors have
invoked this timescale comparison to describe how the day-night
temperature differences should depend on pressure, atmospheric
opacity, stellar irradiation, and other factors
\citep[e.g.,][]{Cooper:2005,
  showman_2008, Showmanetal_2009, fortney-etal-2008,
  Lewis:2010, Rauscher:2010, Cowan_2011,Menou:2012fu,perna_2012,Ginzburg:2015a}.
  
\indent One would expect that
hot Jupiters have short advective timescales due to their fast zonal
winds. This is observationally evident from phase curves from tidally-locked planets
with $\tau_{\mathrm{adv}} \sim \tau_{\mathrm{rad}}$, which
consistently show a peak in brightness just before secondary
eclipse. This indicates that the point of highest emitted flux (``hot
spot'') is eastward of the substellar point (the point of peak
absorbed flux), due to downwind advection from a superrotating\footnote{Superrotation occurs where the zonal-mean atmospheric circulation of a planet has a greater angular momentum per unit mass than the planet itself at the equator.} equatorial jet \citep{Showman_Polvani_2011}. The full-phase observations of HD 189733b
\citep{Knutson_2007}, HD 209458b \citep{Zellem:2014}, and WASP-43b
\citep{Stevenson:2014} show hot spot offsets. These offsets agree with
those predicted from corresponding circulation models
\citep{Showmanetal_2009,Kataria:2014}. Hence, we have observational
confirmation that hot Jupiters have fast ($\sim$ kilometers/second)
zonal winds.

\indent As discussed above, fast zonal winds are a robust feature of
hot Jupiter general circulation models (GCMs), and notably also exist when the model hot Jupiter has a large eccentricity
\citep{Lewis:2010,kataria_2013}. However, these circulation models
show a range of dayside-nightside temperature
differences, showing no clear trend with wind speeds. \cite{perna_2012} examined how heat redistribution is
affected by incident stellar flux, showing that the nightside/dayside
flux ratio decreases with increasing incident stellar flux. This trend
is akin to the observational trend shown in \Fig{fig:Aobs}. This trend
was explained in \cite{perna_2012} by calculating the ratio of
$\tau_{\mathrm{adv}}/\tau_{\mathrm{rad}}$, which increases with
incident stellar flux in their models. 

\indent As pointed out by \citet{Perez-Becker:2013fv}, there exist several
issues with the idea that a timescale comparison between
$\tau_\mathrm{adv}$ and $\tau_\mathrm{rad}$ governs the amplitude of
the day-night temperature differences.  First, although it is
physically motivated, this timescale comparison has never been derived
rigorously from the equations of motion; as such, it has always
constituted an ad-hoc albeit plausible hypothesis, as opposed to a
theoretical result.  Second, the comparison between
$\tau_{\mathrm{adv}}$ and $\tau_{\mathrm{rad}}$ does not include any
obvious role for other timescales that are important, including those
for planetary rotation, horizontal and vertical wave propagation, and
frictional drag (if any).  These processes influence the circulation
and thus one might expect the timescale comparison to depend on them.
Third, the comparison is not predictive---$\tau_\mathrm{adv}$ depends
on the horizontal wind speeds, which are only known {\it a posteriori}.
Hence, it is only possible to evaluate the comparison between
advective and radiative timescales if one already has a numerical
model (or theory) for the atmospheric circulation, which is necessarily related to other relevant timescales governing the circulation.

 \indent To show how other timescales play a large role in determining the
atmospheric circulation, consider the equatorial regions of Earth.  On
Earth, horizontal temperature gradients in the tropics are weak, and
the radiative cooling to space that occurs in Earth's troposphere is
balanced primarily by vertical advection rather than horizontal
advection---a balance known as the weak temperature gradient (WTG)
regime \citep{Polvani:2001, Sobel:2001, Sobel:2002}. This vertical advection timescale is related to the efficacy of lateral wave propagation. When gravity, Rossby, or Kelvin waves propagate, they induce vertical motion, which locally advects the air parcels upward or downward. If these waves are able to propagate away, and rotation plays only a modest role, then this wave-adjustment tends to leave behind a state with flat isentropes---which is equivalent to erasing the horizontal temperature differences \citep{Bretherton:1989,Showman_2013_terrestrial_review}. Given
that adjustment of isentropes due to propagating waves is known to occur on planetary
scales on both Earth \citep{Matsuno:1966,Gill:1980} and exoplanets
\citep{Showman:2010,Showman_Polvani_2011,Tsai:2014},
\cite{Perez-Becker:2013fv} suggested that wave
adjustment likewise acts to lessen horizontal temperature differences in hot
Jupiter atmospheres.

\indent In hot Jupiter atmospheres, planetary-scale Kelvin and Rossby waves
are generated by the large gradient in radiative heating from dayside
to nightside \citep{Showman_Polvani_2011}. Unlike in Earth's atmosphere, these waves exhibit a meridional half-width stretching nearly from equator to pole, as the Rossby deformation radius is approximately equal to the planetary radius \citep{showman_2002}.  These waves cause horizontal convergence/divergence that
forces vertical motion through mass continuity. This vertical motion
moves isentropes vertically, and if the waves are not damped this
leads to a final state with flat isentropes.  However, if these Kelvin
and Rossby waves cannot propagate (i.e. are damped), then this process
cannot occur. Hence, the ability of wave adjustment processes to
lessen horizontal temperature gradients can be weakened by damping of
propagating waves. As shown in \Fig{fig:Aobs}, damping processes that
increase day-night temperature differences seem to increase in
efficacy with increasing equilibrium temperature. The most natural
damping process is radiative cooling, which should increase in efficiency with the cube of
equilibrium temperature \citep{showman_2002}. Additionally, frictional drag on
the atmosphere can reduce the ability of wave adjustment to reduce
longitudinal temperature gradients. This drag could either be due to
turbulence \citep{Li:2010,Youdin_2010} or the Lorentz force in a
partially ionized atmosphere threaded by a dipole magnetic field
\citep{Batygin_2010,Perna_2010_1,Menou:2012fu,batygin_2013,Rauscher_2013,Rogers:2020,Rogers:2014}. Both
of these processes should increase fractional day-night temperature
differences with increasing equilibrium temperature, helping explain
\Fig{fig:Aobs}. However, it is not obvious a priori whether radiative effects
or drag should more efficiently damp wave adjustment in hot Jupiter
atmospheres. 

\indent To understand the mechanisms controlling day-night temperature differences---including their dependence on radiative and frictional effects---\cite{Perez-Becker:2013fv} introduced a shallow-water model with a single active layer representing the atmosphere, which overlies a deeper layer, representing the interior, whose dynamics are fixed and assumed to be quiescent.  \cite{Perez-Becker:2013fv} performed numerical simulations over a broad range of drag and radiative timescales, which generally showed that strong radiation and frictional drag tend to promote larger day-night temperature differences. They then compared their model results to derived analytic theory and found good agreement between the expected fractional dayside-nightside temperature differences and model results, albeit with minor effects not captured in the theory. This theory showed rigorously that wave adjustment allows for reduced horizontal temperature differences in hot Jupiter atmospheres. They also showed that horizontal wave propagation is mainly damped by radiative effects, with potential drag playing a secondary, but crucial, role. As a result, they found that the strength of radiative heating/cooling is the main governor of dayside-nightside temperature differences in hot Jupiter atmospheres. Nevertheless, because the model is essentially two-dimensional (in longitude and latitude) and lacks a vertical coordinate, questions exist about how the results would carry over to a fully three-dimensional atmosphere.

\indent Here we extend the work of \cite{Perez-Becker:2013fv} to fully three-dimensional atmospheres. The use of the full three-dimensional primitive equations enables us to present a predictive analytic understanding of dayside-nightside temperature differences and wind speeds that can be directly compared to observable quantities. Our analytic theory is accompanied by numerical models which span a greater range of radiative forcing and drag parameter space than \cite{Perez-Becker:2013fv}, enabling quantitative validation of these analytic results. We keep the radiative forcing simple in order to promote a physical understanding.  Despite this simplification, we emphasize that this is the first fully predictive analytic theory for the day-night temperature differences of hot Jupiter atmospheres in three dimensions. \\
\indent This paper is organized as follows. In \Sec{sec:methods}, we describe our methods, model setup, and parameter space explored. We discuss the results of our numerical parameter study of dayside-nightside temperature differences in \Sec{sec:numerical}. In \Sec{sec:theory}, we develop our theory in order to facilitate a comparison to numerical results in \Sec{sec:results}. In \Sec{sec:discussion}, we explore the implications of our model results in the context of previous observations and theoretical work, and express conclusions in \Sec{sec:conclusions}.

\section{Model}
\label{sec:methods}
We adopt the same physical model for both our numerical and
  analytic solutions.  GCMs with accurate radiative transfer have
  proven essential for detailed comparison with observations
  \citep{Showmanetal_2009, showman_2013_doppler, Kataria:2014};
  however, our goal here is to promote analytic tractability and a
  clean environment in which to understand dynamical mechanisms, and so
  we drive the circulation using a simplified 
  Newtonian heating/cooling scheme (e.g. \citealp{showman_2002,Cooper:2005}). This enables us to
systematically vary the dayside-nightside thermal forcing and control
the rate at which temperature relaxes to a fixed radiative equilibrium
profile. We also incorporate a drag term in the equations to
investigate how day-night temperature differences are modified by the
combined effects of atmospheric friction and differential stellar
irradiation. Our model setup is nearly identical to that in \cite{Liu:2013}.

\subsection{Dynamical equations}
We solve the horizontal momentum, vertical momentum, continuity, energy equation, and ideal gas equation of state (i.e. the hydrostatic primitive equations), which, in pressure coordinates, are:
\begin{equation}
\label{eq:1}
\frac{d {\bf v}}{dt} + f \hat{\bf k} \times {\bf v} + \nabla \Phi = \mathcal{F}_{\mathrm{drag}} + \mathcal{D}_{\mathrm{S}}  \mathrm{,}
\end{equation}
\begin{equation}
\frac{\partial \Phi}{\partial p} + \frac{1}{\rho} = 0 \mathrm{,}
\end{equation}
\begin{equation}
\nabla \cdot {\bf v} + \frac{\partial \omega}{\partial p} = 0 \mathrm{,}
\end{equation}
\begin{equation}
T\frac{d(\mathrm{ln}\theta)}{dt} = \frac{dT}{dt} - \frac{\omega}{\rho c_{p}} = \frac{q}{c_{p}} + \mathcal{E}_{\mathrm{S}} \mathrm{,}
\label{eq:heating}
\end{equation}
\begin{equation}
\label{eq:4}
p = \rho R T\mathrm{.}
\end{equation}
We use the following symbols: pressure $p$, density $\rho$,
temperature $T$, specific heat at constant pressure $c_{p}$, specific
gas constant $R$, potential temperature\footnote{Potential temperature
  is defined as $\theta = T (p/p_0)^{\kappa}$, where $\kappa =
  R/c_p$, which is here assumed constant. The potential
  temperature is the temperature that an air parcel would have if
  brought adiabatically to an atmospheric reference pressure $p_0$. We
  choose a reference pressure of $p_0 = 1$ bar, but the solution is
  independent of the value of $p_0$ chosen.} $\theta$, horizontal
velocity (on isobars) ${\bf v}$, horizontal gradient on isobars
$\nabla$, vertical velocity in pressure coordinates $\omega = dp/dt$,
geopotential $\Phi = gz$, Coriolis parameter $f = 2\Omega
\mathrm{sin}\phi$ (with $\Omega$ planetary rotation rate, here
equivalent to orbital angular frequency, and $\phi$ latitude), and
specific heating rate $q$. In this coordinate system, the total
(material) derivative is $d/dt = \partial/\partial t + {\bf v} \cdot
\nabla + \omega \partial/\partial p$.  $\mathcal{F}_{\mathrm{drag}}$
represents a drag term that we use to represent missing physics
\citep{Rauscher:2012}, for example drag due to turbulent mixing
\citep{Li:2010,Youdin_2010}, or the Lorentz force
\citep{Perna_2010_1,Rauscher_2013}. The terms
  $\mathcal{D}_{\mathrm{S}}$ and $\mathcal{E}_{\mathrm{S}}$ represent
  a standard fourth-order Shapiro filter, which smooths grid-scale 
  variations while minimally affecting the flow at larger scales,
  and thereby helps to maintain numerical stability in our numerical
  integrations. Because the Shapiro filter terms do not affect the global structure of our equilibrated numerical solutions, they are negligible in comparison to the other terms in the equations at the near-global
  scales captured in our analytic theory. As a result, we neglect them from our
  analytic solutions.
 \begin{figure*}
	\centering
	\includegraphics[width=.9\textwidth]{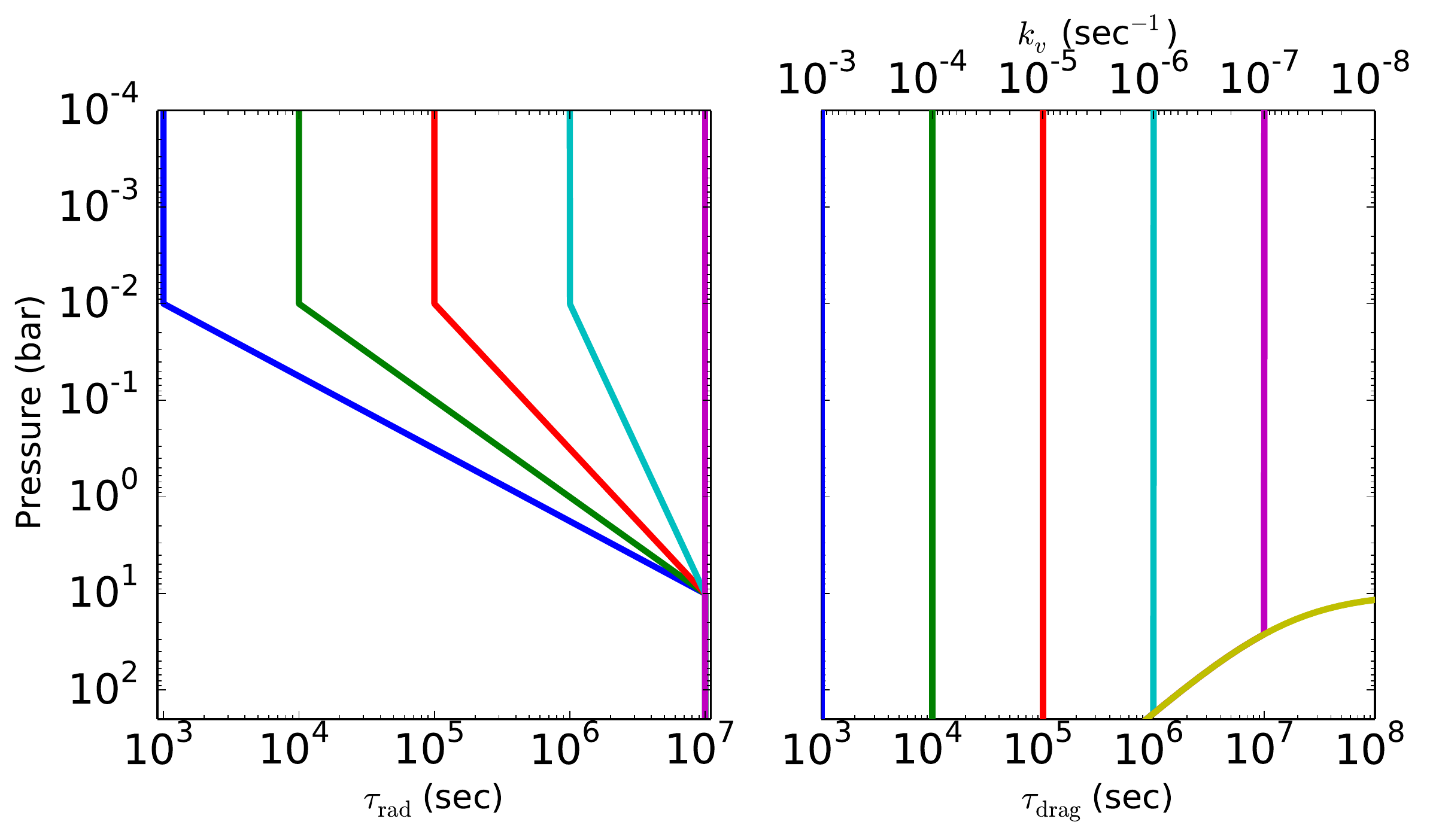}
	\caption{Radiative forcing and drag profiles used in the numerical model. Left: Radiative timescale vs. pressure, for all assumed $\tau_{\mathrm{rad,top}} = 10^3 - 10^7$ sec. Right: Drag timescale as a function of pressure, for each of assumed spatially constant $\tau_{\mathrm{drag}} = 10^3 - \infty$ sec. The drag constant $k_v = \tau_{\mathrm{drag}}^{-1}$ is shown on the upper x-axis.}
	\label{fig:timescales}
\end{figure*}
\subsection{Thermal forcing and frictional drag}
\indent We represent the radiative heating and cooling using a Newtonian heating/cooling scheme, which relaxes the temperature toward a prescribed radiative equilibrium temperature, $T_{\mathrm{eq}}$, over a specified radiative timescale $\tau_{\mathrm{rad}}$:
\begin{equation}
\frac{q}{c_{p}} =  \frac{T_{\mathrm{eq}}(\lambda, \phi, p)-T(\lambda, \phi, p, t)}
{\tau_{\mathrm{rad}}(p)} \mathrm{.}
\end{equation}
In this scheme, $T_{\mathrm{eq}}$ depends on longitude $\lambda$, latitude $\phi$, and pressure, while, for simplicity, $\tau_{\mathrm{rad}}$ varies only with pressure. The radiative equilibrium profile is set to be hot on the dayside and cold on the nightside:
\begin{equation}
\everymath=\expandafter{\displaystyle}
T_{\mathrm{eq}} \left(\lambda,\phi,p\right) = \begin{cases} T_{\mathrm{night,eq}}(p) + \Delta T_{\mathrm{eq}}(p) \mathrm{cos} \lambda \mathrm{cos} \phi \ \mathrm{dayside,}\\ T_{\mathrm{night,eq}}(p) \ \ \ \ \ \ \ \ \ \ \ \ \ \ \ \ \ \ \ \ \ \ \ \ \ \  \mathrm{nightside.}\end{cases}
\end{equation}
Here $\lambda$ is longitude, $T_{\mathrm{night,eq}}(p)$ is the radiative equilibrium temperature profile on the nightside and $T_{\mathrm{night,eq}}(p) + \Delta T_{\mathrm{eq}}(p)$ is that at the substellar point. To acquire the nightside heating profile, $T_{\mathrm{night,eq}}$, we take the temperature profile of HD 209458b from \cite{Iro:2005} and subtract our chosen $\Delta T_{\mathrm{eq}}(p)/2$. We specify $\Delta T_{\mathrm{eq}}$ as in \cite{Liu:2013}, setting it to be a constant $\Delta T_{\rm eq,top}$ at pressures less than $p_{\rm eq,top}$, zero at pressures greater than $p_{\rm bot}$, and varying linearly with log pressure in between:
\begin{equation}
\everymath=\expandafter{\displaystyle}
\Delta T_{\mathrm{eq}}(p) = \begin{cases} \Delta T_{\mathrm{eq,top}} \hspace{3.2cm} p < p_{\mathrm{eq,top}} \\ \Delta T_{\mathrm{eq,top}} \frac{\mathrm{ln}(p/p_{\mathrm{bot}})}{\mathrm{ln}(p_{\mathrm{eq,top}}/p_{\mathrm{bot}})} \hspace{0.8cm} p_{\mathrm{eq,top}} < p < p_{\mathrm{bot}} \\ 0 \hspace{2pt} \mathrm{Kelvin} \hspace{3.2cm} p > p_{\mathrm{bot}} \mathrm{.}\end{cases} 
\end{equation}
As in \cite{Liu:2013}, we assume $p_{\mathrm{eq,top}} = 10^{-3}$ bars and $p_{\mathrm{bot}} = 10$ bars. However, we vary $\Delta T_{\mathrm{eq,top}}$ from $1000 - 0.001$ Kelvin, ranging from highly nonlinear to linear numerical solutions. 

Radiative transfer calculations show that the radiative time
  constant is long at depth and short aloft \citep{Iro:2005,
    Showman:2008}.  To capture this behavior, we adopt the same
  functional form for $\tau_{\mathrm{rad}}$ as \cite{Liu:2013}:
  $\tau_{\mathrm{rad}}$ is set to a large constant $\tau_{\rm
    rad,bot}$ at pressures greater than $p_{\rm bot}$, a 
    (generally smaller) constant $\tau_{\rm rad,top}$ at pressures
    less than $p_{\rm rad,top}$, and varies continuously in between:
\begin{equation}
\label{eq:taurad}
\everymath=\expandafter{\displaystyle}
\tau_{\mathrm{rad}}(p) = \begin{cases} \tau_{\mathrm{rad,top}} \hspace{2cm} p < p_{\mathrm{rad,top}} \\ \tau_{\mathrm{rad,bot}} \left(\frac{p}{p_{\mathrm{bot}}}\right)^\alpha \hspace{.5cm} p_{\mathrm{rad,top}} < p < p_{\mathrm{bot}} \\ \tau_{\mathrm{rad,bot}} \hspace{2cm} p > p_{\mathrm{bot}} \mathrm{,} 	\end{cases}
\end{equation}
with 
\begin{equation}
\alpha = \frac{\mathrm{ln}(\tau_{\mathrm{rad,top}}/ \tau_{\mathrm{rad,bot}})}{\mathrm{ln}(p_{\mathrm{rad,top}}/p_{\mathrm{bot}})} \mathrm{.}
\end{equation}
Here, as in \cite{Liu:2013}, we set $p_{\mathrm{rad,top}} = 10^{-2} \hspace{2pt} \mathrm{bars}$ and $p_{\mathrm{rad,bot}}=10\hspace{2pt}\mathrm{bars}$. Note that the pressures above which $\Delta T_{\mathrm{eq}}$ and $\tau_{\mathrm{rad}}$ are fixed to a constant value at the top of the domain are different, in order to be fully consistent with the model setup of Liu \& Showman (2013). The model is set up such that the circulation forced by Newtonian heating/cooling has three-dimensional temperature and wind distributions that are similar to results from simulations driven by radiative transfer. This motivated the choices of $p_{\mathrm{rad,top}}$ and $p_{\mathrm{eq,top}}$, along with the values of other fixed parameters in the model. \\
\indent The various $\tau_{\mathrm{rad}}$-pressure profiles used in our models (for different assumed $\tau_{\mathrm{rad,top}}$) are shown on the left hand side of \Fig{fig:timescales}.  We choose $\tau_{\mathrm{rad,bot}} = 10^7$ sec, which is long compared to relevant dynamical and rotational timescales but short enough to allow us to readily integrate to equilibrium. For the purposes of our study, we vary $\tau_{\mathrm{rad,top}}$ from $10^3 - 10^7$ sec, corresponding to a range of radiative forcing. 

We introduce a linear drag in the horizontal momentum equation, given by
\begin{equation}
\mathcal{F}_{\mathrm{drag}}  = -k_v(p) {\bf v} \mathrm{,}
\end{equation}
where $k_v(p)$ is a pressure-dependent drag coefficient.  This drag
has two components:

\begin{itemize}
\item First, we wish to examine how forces that crudely parameterize
  Lorentz forces affect the day-night temperature differences. This could be represented with a drag coefficient that
  depends on longitude, latitude, and pressure and, moreover, differs in all three dimensions \cite[e.g.][]{Perna_2010_1,
    Rauscher_2013}. However, the Lorentz force should depend strongly on the ionization fraction and therefore the local temperature, requiring full numerical magnetohydrodynamic solutions to examine the effects of magnetic ``drag'' in detail, e.g. \cite{batygin_2013,Rogers:2014,Rogers:2020}. For simplicity and analytic
  tractability, we represent this component with a spatially constant
  drag timescale $\tau_{\rm drag}$, corresponding to a drag
  coefficient $\tau_{\rm drag}^{-1}$.  We systematically explore
  $\tau_{\rm drag}$ values (in sec) of $10^3$, $10^4$, $10^5$, $10^6$,
  $10^7$, and $\infty$.  The latter corresponds to the drag-free limit.
  Such a scheme was already explored by \cite{showman_2013_doppler}.

\item Second, following \cite{Liu:2013}, we introduce a ``basal'' drag
  at the bottom of the domain, which crudely parameterizes
  interactions between the vigorous atmospheric circulation and a
  relatively quiescent planetary interior.  For this component, the
  drag coefficient is zero at pressures less than
  $p_{\mathrm{drag,top}}$ and is
  $\tau_{\mathrm{drag,bot}}^{-1}(p-p_{\mathrm{drag,top}})/
  (p_{\mathrm{drag,bot}}-p_{\mathrm{drag,top}})$ at pressures greater
  than $p_{\mathrm{drag,top}}$, where $p_{\rm drag,bot}$ is the mean
  pressure at the bottom of the domain (200 bars) and $p_{\rm
    drag,top}$ is the lowest pressure where this basal drag component
  is applied.  Thus, the drag coefficient varies from $\tau_{\rm
    drag,bot}^{-1}$ at the bottom of the domain to zero at a pressure
  $p_{\rm drag,top}$; this scheme is similar to that in \cite{Held:1994}.
We take $p_{\rm drag,top}=10\,$bars, and set
  $\tau_{\rm drag,bot}=10\,$days.  Thus, basal drag acts only at
  pressures greater than 10 bars and has a minimum characteristic
  timescale of 10 days at the bottom of the domain, increasing to
  infinity (meaning zero drag) at pressures less than 10 bars. We emphasize that the precise value is not critical.  Changing the drag time constant at the base to 100 days, for example (corresponding to weaker drag) would lead to slightly faster wind speeds at the base of the model, and would require longer integration times to reach equilibrium, but would not qualitatively change our results.
\end{itemize}
To combine the two drag schemes, we simply set the drag coefficient to
be the smaller of the two individual drag coefficients at each
individual pressure, leading to a final functional form for the drag
coefficient:
\begin{equation}
k_v(p) = \max\left[\tau_{\rm drag}^{-1},
  \tau_{\mathrm{drag,bot}}^{-1}\frac{(p-p_{\mathrm{drag,top}})}
       {(p_{\mathrm{drag,bot}}-p_{\mathrm{drag,top}})}\right]
\end{equation}
The righthand side of \Fig{fig:timescales} shows the various $\tau_{\mathrm{drag}}(p)$ profiles used in our models. Corresponding values for $k_v(p)$ are given along the top axis.

We adopt planetary parameters ($c_{p}$, $R$, $\Omega$, $g$, $R$) relevant for HD 209458b. This includes specific heat $c_p = 1.3 \times 10^4 \hspace{2pt} \mathrm{J} \hspace{2pt} \mathrm{kg}^{-1}\mathrm{K}^{-1}$, specific gas constant $R = 3700 \hspace{2pt} \mathrm{J}  \hspace{2pt}\mathrm{kg}^{-1}\mathrm{K}^{-1}$, rotation rate $\Omega = 2.078 \times 10^{-5}\hspace{2pt} \mathrm{s}^{-1}$, gravity $g = 9.36 \hspace{2pt} \mathrm{m} \hspace{2pt} \mathrm{s}^{-2}$, and planetary radius $a = 9.437  \times 10^{7} \hspace{2pt} \mathrm{m}$. Though we use parameters relevant for a given hot Jupiter, our results are not sensitive to the precise values used. Moreover, we emphasize that the qualitative model behavior should not be overly sensitive to numerical parameters such as the precise values of $p_{\mathrm{rad,top}}$, $p_{\mathrm{eq,top}}$, $p_{\mathrm{bot}}$, and so on. Modifying the values of these parameters over some reasonable range will change the precise details of the height-dependence of the day-night temperature difference and wind speeds but will not change the \textit{qualitative} behavior or the dynamical mechanisms we seek to uncover. We would find similar behavior regardless of the specific parameters used, as long as they are appropriate for a typical hot Jupiter. 

\subsection{Numerical details}

Our numerical integrations are performed using the MITgcm
\citep{Adcroft:2004} to solve the equations described above on a
cubed-sphere grid. The horizontal resolution is C32, which is roughly
equal to a global resolution of $128 \times 64$ in longitude and
latitude. There are 40 vertical levels, with the bottom 39 levels
evenly spaced in log-pressure between 0.2 mbars and 200 bars, and a
top layer that extends from 0.2 mbars to zero pressure. Models
  performed at resolutions as high as C128 (corresponding to a global
  resolution of $512\times256$) by \cite{Liu:2013} behave very similar
  to their C32 counterparts, indicating that C32 is sufficient for
  current purposes.  All models are integrated to statistical
  equilibrium. We integrate the model from a state of rest with the temperatures set to the \cite{Iro:2005} temperature-pressure profile.  Note that this system does not exhibit sensitivity to initial conditions \citep{Liu:2013}. For the most weakly nonlinear runs described in \Sec{sec:param}, reaching equilibration required $25,000$ Earth days of model integration time. However, our full grid of simulations varying radiative and drag timescales with a fixed equilibrium day-night temperature difference required $\lesssim 5,000$ days of integration.

\section{Numerical Results}
\label{sec:numerical}
\begin{figure}
	\includegraphics[width=0.475\textwidth]{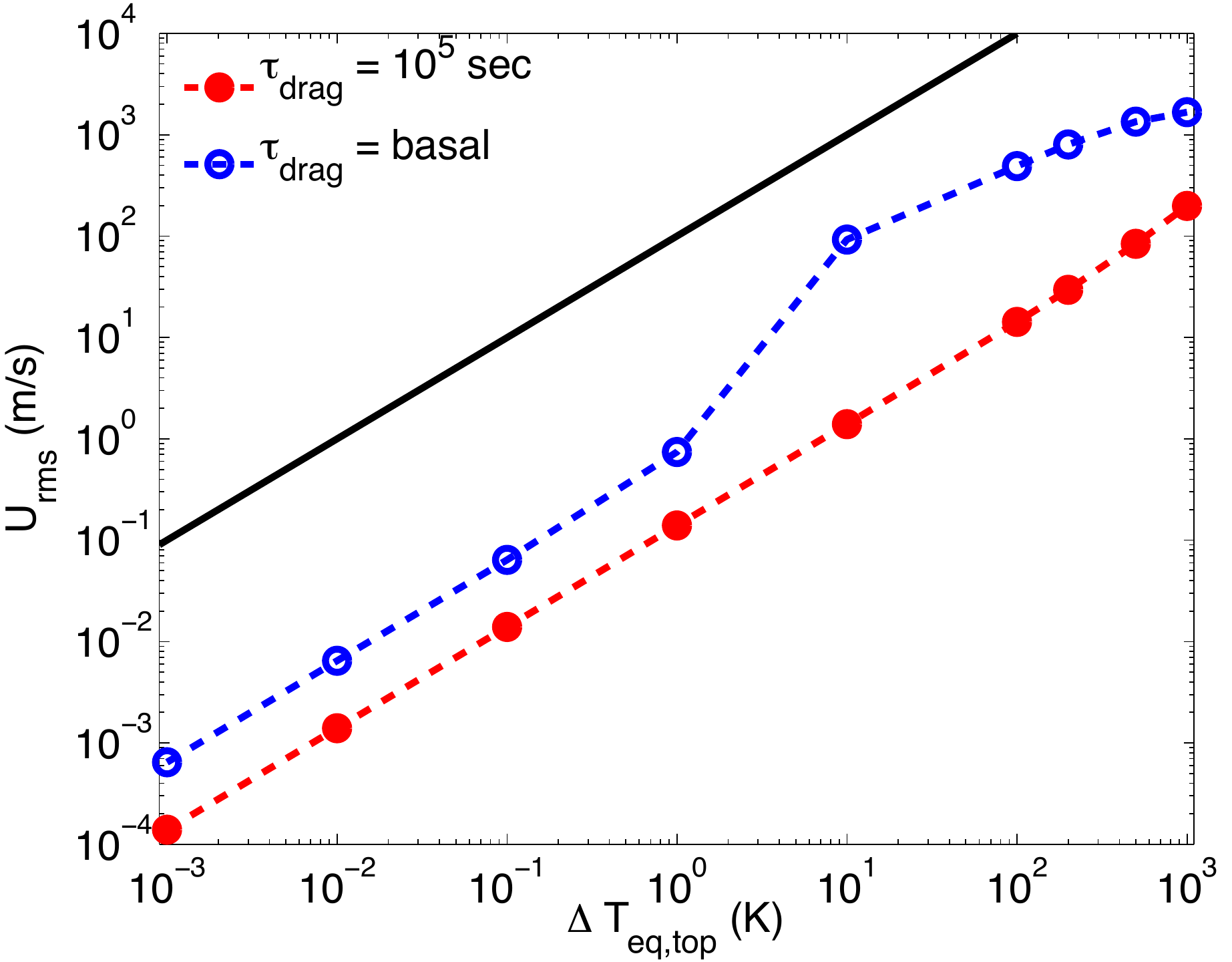}
	\caption{Root-mean-square (RMS) horizontal wind speed at a pressure of $80 \hspace{2pt} \mathrm{mbars}$ plotted against equilibrium dayside-nightside temperature differences $\Delta T_{\mathrm{eq,top}}$. These day-night temperature differences set the forcing amplitude of the circulation. When $\Delta T_{\mathrm{eq,top}}$ is small, the RMS velocities respond linearly to forcing, and when $\Delta T_{\mathrm{eq,top}}$ is large, the RMS velocities respond nonlinearly. The transition from nonlinear to linear response occurs at different $\Delta T_{\mathrm{eq,top}}$ depending on whether or not spatially constant drag is applied. When there is not spatially constant drag (blue open circles) and only basal drag is applied, the transition occurs at $\Delta T_{\mathrm{eq,top}} \sim 0.1 \hspace{2pt} \mathrm{Kelvin}$. When $\tau_{\mathrm{drag}} = 10^5 \hspace{2pt} \mathrm{sec}$ (red filled circles), the transition occurs at $\Delta T_{\mathrm{eq,top}} \sim 10 \hspace{2pt} \mathrm{Kelvin}$. For comparison, a linear relationship between $U_{\mathrm{rms}}$ and $\Delta T_{\mathrm{eq,top}}$ is shown by the black line.}
	\label{fig:transurms}
\end{figure}
\begin{figure*}
	\includegraphics[height=.7\textheight]{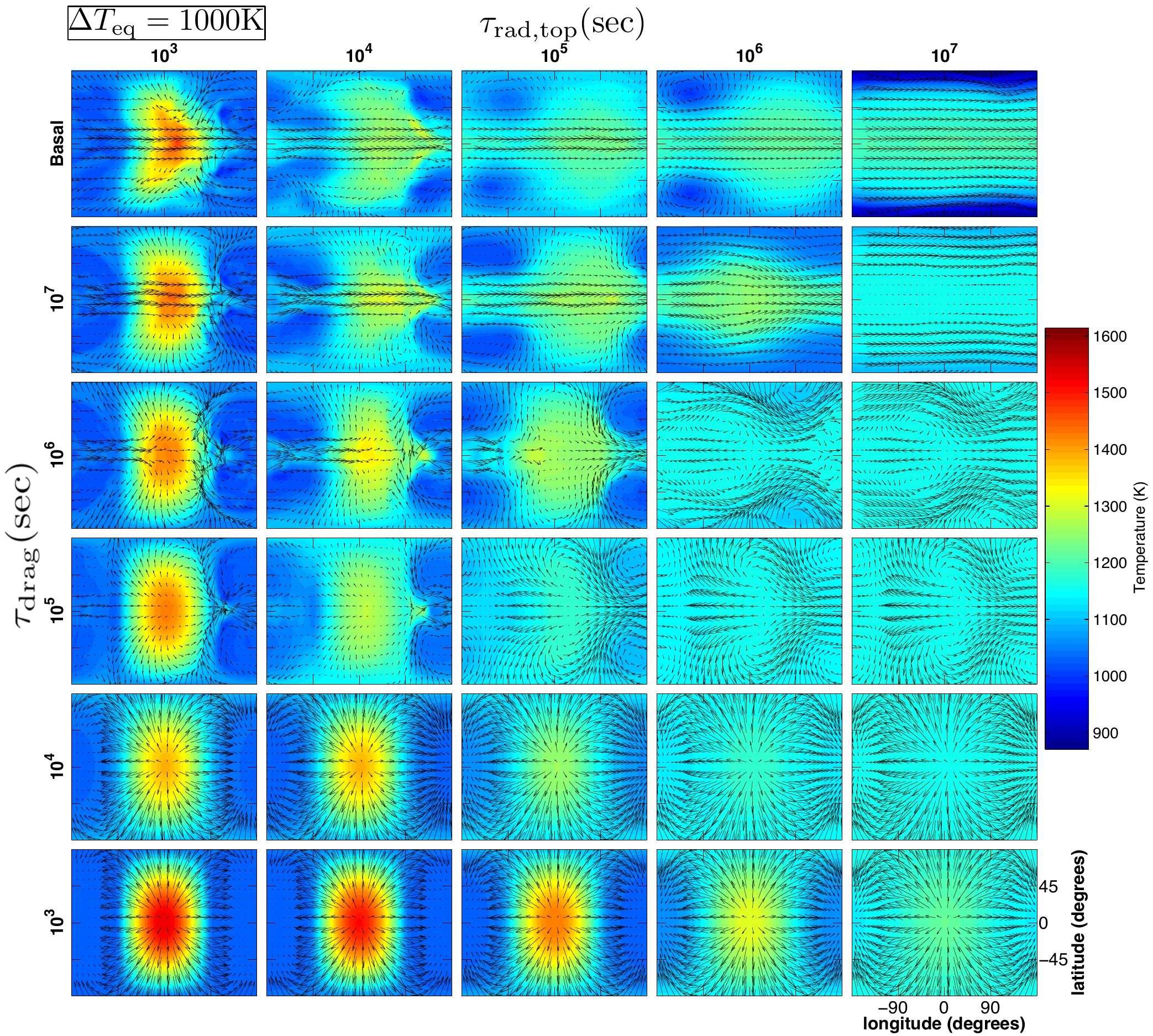}
	\caption{Maps of temperature (colors) and wind (vectors) for suite of 30 GCM simulations varying $\tau_{\mathrm{rad,top}}$ and $\tau_{\mathrm{drag}}$ with $\Delta T_{\mathrm{eq,top}} = 1000$ Kelvin. All maps are taken from the $80$-mbar statistical steady-state end point of an individual model run. All plots share a color scheme for temperature but have independent overplotting of horizontal wind vectors. The substellar point is located at $0^{\circ}, 0^{\circ}$ in each plot, with the lower right plot displaying latitude \& longitude axes.}
	\label{fig:nltempgrid}
\end{figure*}
\begin{figure*}
	\includegraphics[height=.7\textheight]{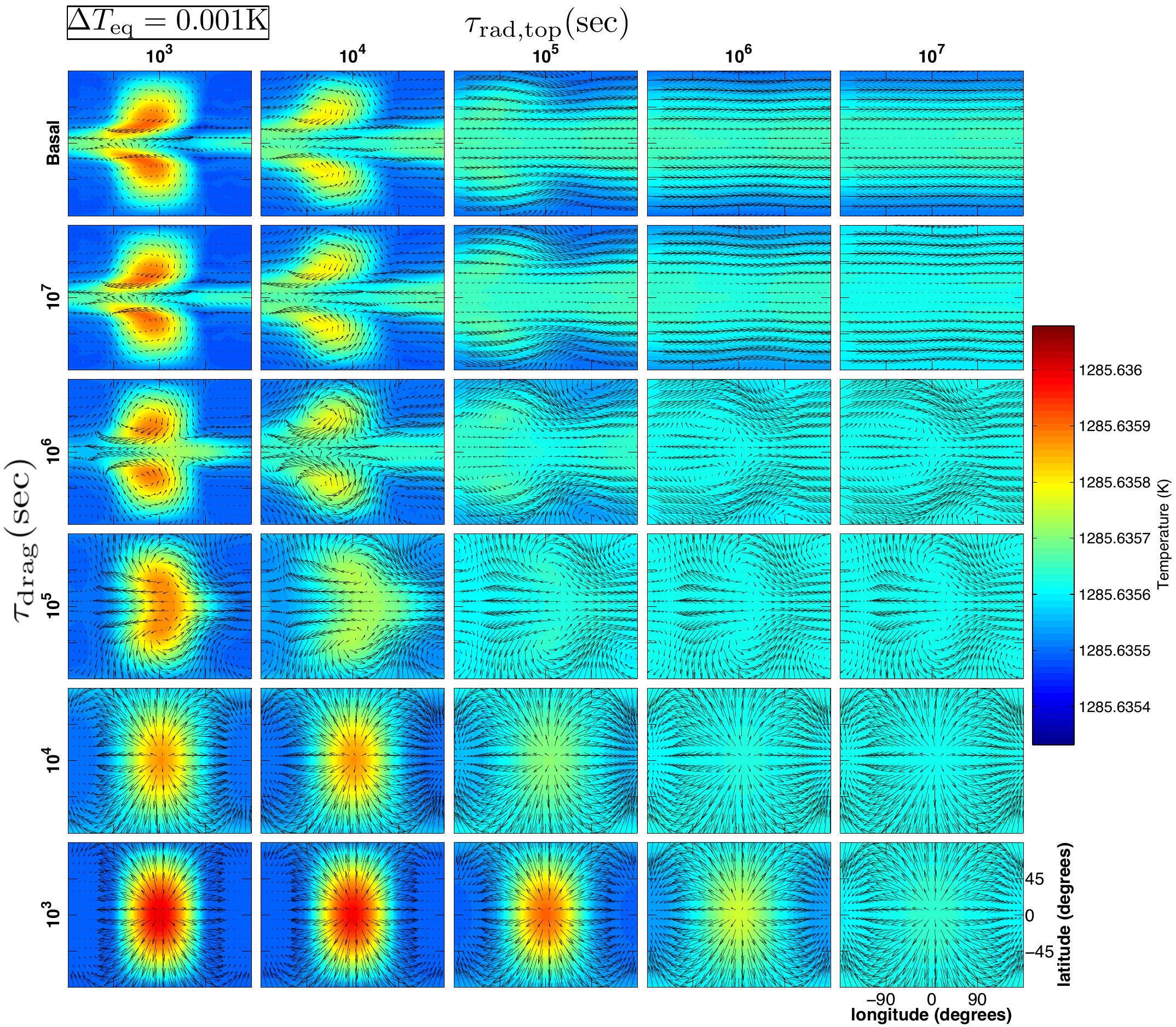}
	\caption{Same as Figure \ref{fig:nltempgrid}, except with $\Delta T_{\mathrm{eq,top}} = 0.001$ Kelvin.}
	\label{fig:lintempgrid}
\end{figure*}
\subsection{Parameter space exploration}
\label{sec:param}
\indent Given the forcing and drag prescriptions specified in
  Section~\ref{sec:methods} and the planetary parameters for a typical
  hot Jupiter, the problem we investigate is one governed by three
  parameters---$\Delta T_{\rm eq,top}$, $\tau_{\rm rad,top}$, and
  $\tau_{\rm drag}$.  Our goal is to thoroughly explore a broad,
  two-dimensional grid of GCM simulations varying $\tau_{\rm rad,top}$ and
  $\tau_{\rm drag}$ over a wide range. Here we first explore the role
  of $\Delta T_{\rm eq}$ so that we may make appropriate choices about
  the values of $\Delta T_{\rm eq}$ to use in the full grid.


\indent The day-night radiative-equilibrium temperature contrast $\Delta
T_{\rm eq}$ represents the amplitude of the imposed radiative
forcing, and controls the amplitude of the resulting flow.  In the
low-amplitude limit ($\Delta T_{\mathrm{eq}} \rightarrow 0$), the
  wind speeds and temperature perturbations are weak, and the
  nonlinear terms in the dynamical equations should become small
  compared to the linear terms.  Thus, in this limit, the solutions
  should behave in a mathematically linear\footnote{By this we mean
    that as $\Delta T_{\rm eq}\to 0$, the solutions of the full
    nonlinear problem should converge toward the mathematical solutions to
    versions of Equations~(\ref{eq:1})--(\ref{eq:heating}) that are
    linearized around a state with zero wind and the background $T(p)$
    profile.} manner: the spatial structure of the
  circulation should become independent of forcing amplitude, and the
  amplitude of the circulation---that is, the wind speeds and
  day-night temperature differences---should vary linearly with
  forcing amplitude.  On the other hand, at very high forcing 
amplitudes (large $\Delta T_{\rm eq}$), the wind speeds and temperature
differences are large, and the solutions behave nonlinearly.

\indent  Therefore, we first performed a parameter sweep of $\Delta
 T_{\mathrm{eq}}$ to understand the transition between linear and
 nonlinear forcing regimes, and to determine the value of $\Delta
   T_{\rm eq}$ at which this transition occurs.  For this parameter
 sweep, we performed a sequence of models varying $\Delta
 T_{\mathrm{eq,top}}$ from $1000$ to $0.001 \hspace{2pt}
 \mathrm{Kelvin}$\footnote{Specifically, we tested values of $\Delta
   T_{\mathrm{eq,top}} = 1000, 500, 200, 100, 10, 1, 0.1, 0.01,$ and
   $0.001\hspace{1pt} \mathrm{K}$.}. We did
 one such sweep using $\tau_{\rm rad,top}=10^4\rm\,s$ and $\tau_{\rm
   drag}=\infty$ (meaning basal-drag only), and another such sweep
 using $\tau_{\rm rad,top}=10^4\rm\,s$ and $\tau_{\rm
   drag}=10^5\rm\,s$.   These sweeps verify that we are indeed in the
 linear limit (where variables such as wind speed and temperature
 respond linearly to forcing) at $\Delta T_{\mathrm{eq,top}}
 \lesssim 0.1$ Kelvin for any $\tau_{\mathrm{drag}}$. Figure
 \ref{fig:transurms} shows how the root-mean-square (RMS) horizontal
 wind speed varies with $\Delta T_{\mathrm{eq,top}}$ for these two
 parameter sweeps. Here, the RMS horizontal wind speed,
 $U_{\mathrm{rms}}$, is defined at a given pressure as:
\begin{equation}
\label{eq:urms}
U_{\mathrm{rms}}(p) = \sqrt{\frac{\int (u^2 + v^2) dA}{A}} \mathrm{.}
\end{equation}
Here the integral is taken over the globe, with $A$ the horizontal area of the globe and $u,v$ the zonal and meridional velocities at a given pressure level, respectively. It is notable that the linear limit is reached at different $\Delta T_{\mathrm{eq,top}}$ values depending on the strength of the drag applied. With a spatially constant $\tau_{\mathrm{drag}} = 10^5$ sec, the models respond linearly to forcing at $\Delta T_{\mathrm{eq,top}} \lesssim 10 \hspace{2pt} \mathrm{Kelvin}$, and are very nearly linear throughout the range of $\Delta T_{\mathrm{eq,top}}$ considered. This causes the $U_{\mathrm{rms}}$-$\Delta T_{\mathrm{eq,top}}$ relationship for $\tau_{\mathrm{drag}} = 10^5$ sec to be visually indistinguishable from a linear slope in \Fig{fig:transurms}. However, without a spatially constant drag, the linear limit is not reached until $\Delta T_{\mathrm{eq,top}} \lesssim 0.1 \hspace{2pt} \mathrm{Kelvin}$, and the dynamics are nonlinear for $\Delta T_{\mathrm{eq,top}}\gtrsim \,1 \rm K$.

The main grid presented involves a parameter study mimicking that of \cite{Perez-Becker:2013fv}, but using the 3D primitive equations. We do so in order to understand mechanisms behind hot Jupiter dayside-nightside temperature differences with the full system of nonlinear primitive equations. We varied $\tau_{\mathrm{rad,top}}$ from $10^3 - 10^7$ sec and $\tau_{\mathrm{drag}}$ from $10^3 - \infty \hspace{2pt} \mathrm{sec}$ (the range of timescales displayed in Figure \ref{fig:timescales}), extending an order of magnitude lower in $\tau_{\mathrm{drag}}$ than \cite{Perez-Becker:2013fv}. We ran this suite of models for both $\Delta T_{\mathrm{eq,top}} = 1000 \hspace{2pt} \mathrm{Kelvin}$ (nonlinear regime, see Figure \ref{fig:nltempgrid}) and $\Delta T_{\mathrm{eq,top}} = 0.001 \hspace{2pt} \mathrm{Kelvin}$ (linear regime, see Figure \ref{fig:lintempgrid}) to better understand the mechanisms controlling dayside-nightside temperature differences. 

\subsection{Description of atmospheric circulation over a wide range of radiative
and frictional timescales}
\indent Figure \ref{fig:nltempgrid} shows latitude-longitude maps of temperature (with overplotted wind vectors) at a pressure of 80 mbars for the entire suite of models performed at $\Delta T_{\mathrm{eq,top}} = 1000 \hspace{2pt} \mathrm{Kelvin}$. These are the statistically steady-state end points of 30 separate model simulations, with $\tau_{\mathrm{rad,top}}$ varying from $10^3 - 10^7 \hspace{2pt} \mathrm{sec}$ and $\tau_{\mathrm{drag}}$ ranging from $10^3 - \infty \hspace{2pt} \mathrm{sec}$. All plots have the same temperature colorscale for inter-comparison. \\
\indent One can identify distinct regimes in this $\tau_{\mathrm{rad,top}}$ and $\tau_{\mathrm{drag}}$ space. First, there is the nominal hot Jupiter regime with $\tau_{\mathrm{drag}} = \infty$ and $\tau_{\mathrm{rad,top}} \lesssim10^5 \hspace{2pt} \mathrm{sec}$, which has been studied extensively in previous work. This regime has a strong zonal jet which manifests as equatorial superrotation. However, there is no zonal jet when $\tau_{\mathrm{drag}} \lesssim 10^5 \hspace{2pt} \mathrm{sec}$. Hence, there is a regime transition in the atmospheric circulation at $\tau_{\mathrm{drag}} \sim 10^6 \hspace{2pt} \mathrm{sec}$ between a strong coherent zonal jet and weak or absent zonal jets. Additionally, there are two separate regimes of dayside-nightside temperature differences as $\tau_{\mathrm{rad,top}}$ varies. When $\tau_{\mathrm{rad,top}}$ is short ($10^3-10^4 \hspace{2pt} \mathrm{sec}$), the dayside-nightside temperature differences are large. When $\tau_{\mathrm{rad,top}} \gtrsim 10^6 \hspace{2pt} \mathrm{sec}$, dayside-nightside temperature differences are small on latitude circles. However, $\tau_{\mathrm{rad,top}}$ is not the only control on dayside-nightside temperature differences. If atmospheric friction is strong, with $\tau_{\mathrm{drag}} \lesssim 10^4 \hspace{2pt} \mathrm{sec}$, dayside-nightside temperature differences are large unless $\tau_{\mathrm{rad,top}}$ is extremely long. \\
\indent Equivalent model inter-comparison to \Fig{fig:nltempgrid} for the $\Delta T_{\mathrm{eq,top}} = 0.001 \hspace{2pt} \mathrm{Kelvin}$ case is presented in Figure \ref{fig:lintempgrid}. The same major trends apparent in the $\Delta T_{\mathrm{eq,top}} = 1000 \hspace{2pt} \mathrm{Kelvin}$ results are seen here in the linear limit. That is, $\tau_{\mathrm{rad,top}}$ is still the key parameter control on dayside-nightside temperature differences. If $\tau_{\mathrm{rad,top}}$ is small, dayside-nightside temperature differences are large, and if $\tau_{\mathrm{rad,top}}$ is large, dayside-nightside differences are small. This general trend is modified slightly by atmospheric friction---if $\tau_{\mathrm{drag}} \lesssim 10^4 \hspace{2pt} \mathrm{sec}$, drag plays a role in determining the dayside-nightside temperature differences. 

A key difference between simulations at high and low $\Delta
T_{\mathrm{eq}}$ is that, under conditions of short $\tau_{\rm
    rad}$ and weak drag (upper-left quadrant of
  Figures~\ref{fig:nltempgrid}--\ref{fig:lintempgrid}), the maximum
  temperatures occur on the equator in the nonlinear limit, but they
  occur in midlatitudes in the linear limit. This can be understood
by considering the force balances. Namely, because the Coriolis force
goes to zero at the equator, the only force that can balance the
pressure gradient at the equator is
advection. Advection is a nonlinear term that scales with the square
of wind speed, and hence this force balance is inherently
nonlinear. As a result, the advection term and pressure gradient force
both weaken drastically at the equator in the linear limit, and the
nonlinear balance cannot hold. This causes the equator to be nearly
longitudinally isothermal in the linear limit, rather than
having large day-night temperature differences as in the
nonlinear case. This phenomenon was already noted by \cite{Showman_Polvani_2011}
and \cite{Perez-Becker:2013fv} in one-layer shallow water models,
and Figure~\ref{fig:lintempgrid} represents an extension of it to the
3D system.

Another difference in the temperature maps between the nonlinear and
linear limit is the orientation of the phase tilts in the long
$\tau_{\mathrm{drag}}$, short $\tau_{\mathrm{rad}}$ upper left
quadrant of \Fig{fig:lintempgrid}. These phase tilts are the exact
opposite of those that are needed to drive superrotation.  The linear
dynamics in the weak-drag limit causing these tilts has been examined
in detail by \cite{Showman_Polvani_2011}, see their Appendix C. They showed analytically that in the limit of long $\tau_{\mathrm{drag}}$, the standing Rossby waves develop phase tilts at low latitudes that are northeast-to-southwest in the northern hemisphere and southeast-to-northwest in the southern hemisphere. This is the opposite of the orientation needed to transport eastward momentum to equatorial regions and drive equatorial superrotation. Hence, the upper left
quadrant of \Fig{fig:lintempgrid} shows key distinctions from the same
quadrant in \Fig{fig:nltempgrid}. In the linear limit, there is no
superrotation, and the ratio of characteristic dayside and nightside
temperatures at the equator is much smaller than in the case with
large forcing amplitude. 

\indent Our model grids in Figures~\ref{fig:nltempgrid} and
  \ref{fig:lintempgrid} exhibit a striking resemblance to the
  equivalent grids from the shallow-water models of \citet[][see their
    Figures 3 and 4]{Perez-Becker:2013fv}. This gives us confidence
that the same mechanisms determining the day-night temperature
differences in their one-layer models are at work in the full 3D system. Additionally, by extending $\tau_{\mathrm{drag}}$
one order of magnitude shorter, we have reached the parameter regime where
drag can cause increased day-night temperature differences even when
$\tau_{\mathrm{rad}}$ is extremely long, allowing a more robust comparison
to theory in this limit.

\indent Despite the distinctions between the nonlinear and linear limits discussed above, both grids show similar overall parameter dependences on radiative forcing and frictional drag. As radiative forcing becomes stronger (i.e., $\tau_{\rm rad,top}$ becomes shorter), day-night temperature differences increase, with drag only playing a role if it is extremely strong. Additionally, drag is the key factor to quell the zonal jet. The fact that the same general trends in day-night temperature differences occur in both the nonlinear and linear limit suggests that the same qualitative mechanisms are controlling day-night temperature differences in both cases, although nonlinearities will of course introduce quantitative differences at sufficiently large $\Delta T_{\rm eq}$.  This makes it likely that a simple analytic theory can explain the trends seen in Figures \ref{fig:nltempgrid} and \ref{fig:lintempgrid}. We develop such a theory in \Sec{sec:theory}, and continue in \Sec{sec:results} to compare our results to the numerical solutions presented in this section. 

\section{Theory}
\label{sec:theory}
\subsection{Pressure-dependent theory}
\label{sec:pressdepA}

\indent We seek approximate analytic solutions to the problem
  posed in Sections~\ref{sec:methods}--\ref{sec:numerical}.  Specifically,
  here we present solutions for the pressure-dependent day-night temperature
  difference and the characteristic horizontal and vertical wind speeds
  as a function of the external control parameters ($\Delta T_{\rm eq,top}$,
  $\tau_{\rm rad,top}$, $\tau_{\rm drag}$, and the planetary parameters). In this theory, we do not distinguish variations in longitude from variations in latitude. As a result, we assume that the day-to-night and equator-to-pole temperature differences are comparable. Most GCM studies produce relatively steady hemispheric-mean circulation patterns
  \citep[e.g.,][]{Showmanetal_2009, Liu:2013}, including our own simulations shown in Section \ref{sec:numerical}, and so we seek steady solutions to the primitive equations (\ref{eq:1})--(\ref{eq:4}).  For convenience,
  we here cast these in log-pressure coordinates \citep{Andrews:1987,Holton:2013}:
\begin{equation}
\label{eq:hpecart1}
{\bf v} \cdot \nabla {\bf v} + w^{\star} \frac{\partial {\bf v}}{\partial z^{\star}} + f {\bf k} \times {\bf v} = - \nabla \Phi - \frac{{\bf v}}{\tau_{\mathrm{drag}}} \mathrm{,}
\end{equation}
\begin{equation}
\label{eq:hpelogphydro}
\frac{\partial \Phi}{\partial z^{\star}} = RT \mathrm{,}
\end{equation}
\begin{equation}
\label{eq:hpecart2}
\nabla \cdot {\bf v} + e^{z^\star} \frac{\partial (e^{-z^\star} w^{\star})}{\partial z^{\star}} = 0 \mathrm{,}
\end{equation}
\begin{equation}
\label{eq:hpecart3}
{\bf v} \cdot \nabla T + w^{\star} \frac{N^2 H^2}{R} = \frac{T_{\mathrm{eq}}-T}{\tau_{\mathrm{rad}}} \mathrm{.}
\end{equation}
Equation (\ref{eq:hpecart1}) is the horizontal momentum equation, \Eq{eq:hpelogphydro} the vertical momentum equation (hydrostatic balance), Equation (\ref{eq:hpecart2}) the continuity equation, and Equation (\ref{eq:hpecart3}) the thermodynamic energy equation. In this coordinate system, $z^{\star}$ is defined as
\begin{equation}
z^{\star} \equiv -\mathrm{ln} \frac{p}{p_{\mathrm{00}}} \mathrm{,}
\end{equation}
with $p_{\mathrm{00}}$ a reference pressure, and the vertical velocity $w^{\star} \equiv dz^{\star}/dt$, which has units of scale heights per second (such that $1/w^\star$ is the time needed for air to flow vertically over a scale height). $N$ is the Brunt-Vaisala frequency and $H=RT/g$ is the scale height. This equation set is equivalent to the steady-state version of Equations (\ref{eq:1}-\ref{eq:4}). Note that drag is explicitly set on the horizontal components of velocity. We use the steady-state system here in order to facilitate comparison with our models, which themselves are run to steady-state with kinetic energy equilibration.

Given the set of Equations (\ref{eq:hpecart1})--(\ref{eq:hpecart3})
above, we can now utilize scaling to give approximate solutions for
comparison to both our fully nonlinear (high $\Delta T_{\mathrm{eq}}$)
and linear (low $\Delta T_{\mathrm{eq}}$) numerical solutions. Here, we step systematically through the equations, starting with the continuity equation, then the thermodynamic energy equation, and finally the momentum equations.

\subsubsection{Continuity equation}

First, consider the continuity equation (\ref{eq:hpecart2}).  The
scaling for the first term on the left hand side is subtle. When
the Rossby number $\mathrm{Ro} \gtrsim 1$, we expect that $\nabla
\cdot {\bf v} \sim \mathcal{U}/\mathcal{L}$, where $\mathcal{U}$ is a characteristic horizontal velocity and $\mathcal{L}$ a characteristic lengthscale of the circulation. However, when
$\mathrm{Ro} \lesssim 1$, geostrophy holds and in principle we could have $\nabla \cdot {\bf v}
\ll \mathcal{U}/\mathcal{L}$. For a purely geostrophic flow, $\nabla
\cdot {\bf v} \equiv -\beta v/f$, with $v$ meridional velocity
\citep[see][Eq.~32]{Showman_2009}. On a sphere, $\beta =
2\Omega\mathrm{cos}\phi/a$, and hence $\beta/f =
\mathrm{cot}\phi/a$. As a result, $\nabla \cdot {\bf v} \approx
\mathcal{U} \mathrm{cot}\phi/a$. For hot Jupiters, $\mathcal{L} \sim
a$, and hence it turns out that for geostrophic flow not too close to
the pole that $\nabla \cdot {\bf v} \sim
\mathcal{U}/\mathcal{L}$. Hence, the scaling $\nabla \cdot {\bf v}
\sim \mathcal{U}/\mathcal{L}$ holds throughout the circulation regimes
considered here.

Now consider the second term on the left side of (\ref{eq:hpecart2}).
Expanding out the derivative yields $-w^\star + \partial
w^\star/\partial z^\star$.  The term $\partial w^\star/\partial
z^\star$ scales as $w^\star/\Delta z^\star$, where $\Delta
z^\star$ is the vertical distance (in scale heights) over which
$w^\star$ varies by its own magnitude.  Thus, one could write
\begin{equation}
\frac{\mathcal{U}}{\mathcal{L}} \sim
\max\left[w^\star,\frac{w^\star}{\Delta z^\star} \right].
\end{equation}
Previous GCM studies suggest that $w^\star$ maintains coherency of
values over several scale heights \citep[e.g.][]{parmentier_2013},
suggesting that $\Delta z^\star$ is several (i.e., greater than one),
in which case the first term on the right dominates.  Defining an
alternate characteristic vertical velocity $\mathcal{W} = H w^\star$,
which gives the approximate vertical velocity in $\rm m\,s^{-1}$, the
continuity equation becomes simply
\begin{equation}
\frac{\mathcal{U}}{\mathcal{L}} \sim \frac{\mathcal{W}}{H}.
\label{cont-simple}
\end{equation}

\subsubsection{Thermodynamic energy equation}
\label{sec:thermo}
\begin{figure}
	\centering
	\includegraphics[width=.475\textwidth]{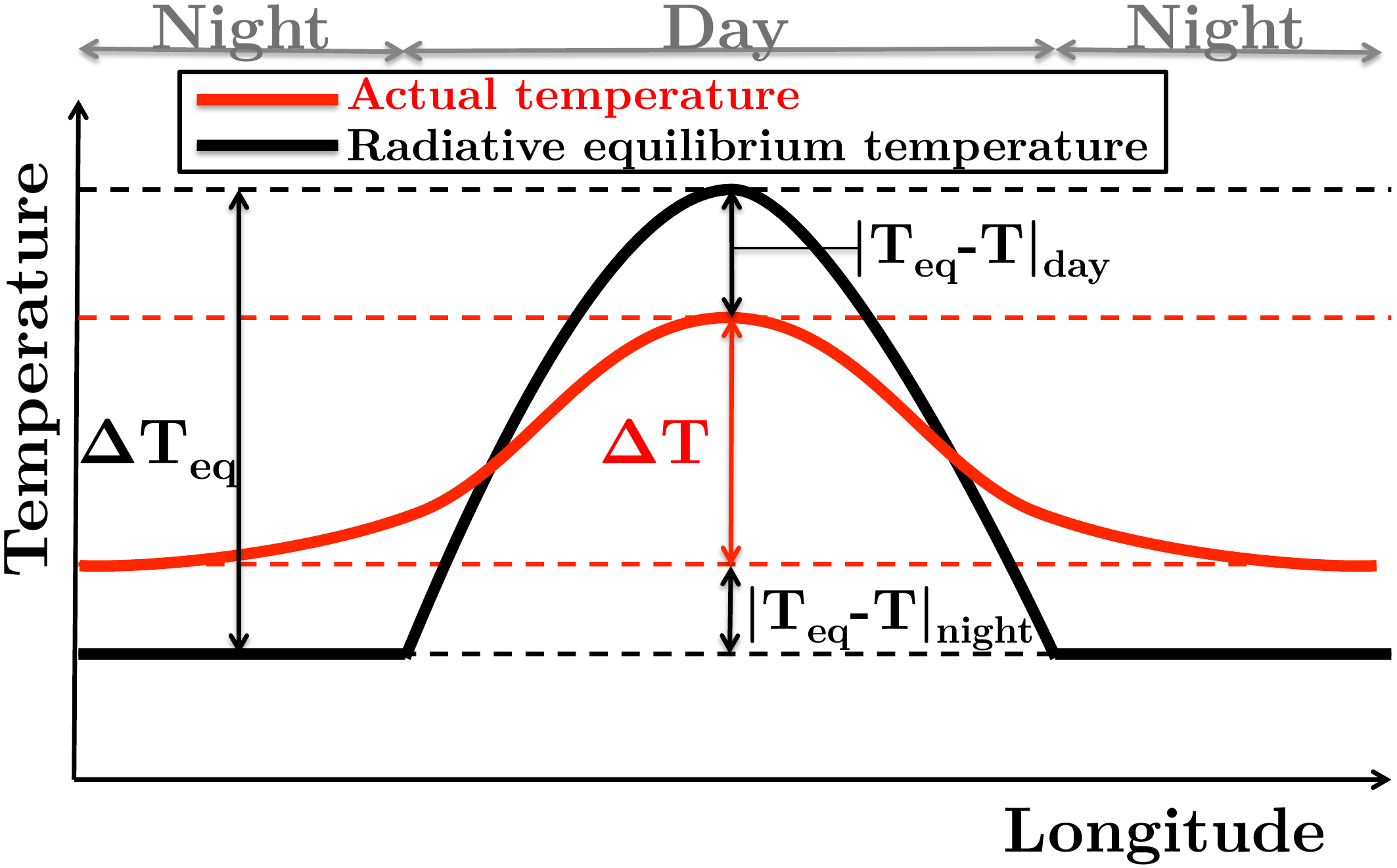}
	\caption{Simplified diagram of the model, schematically displaying longitudinal profiles of actual and radiative equilibrium temperature. We show this schematic to help explain \Eq{eq:deltaT}. The difference between the characteristic actual and radiative equilibrium temperature differences from dayside to nightside, $\Delta T_{\mathrm{eq}} - \Delta T$, is approximately equal to the sum of the characteristic differences on each hemisphere, $\left| T_{\mathrm{eq}} - T \right|_{\mathrm{day}} + \left| T_{\mathrm{eq}} - T \right|_{\mathrm{night}}$.}
	\label{fig:schematic}
\end{figure}

\indent We next consider the thermodynamic energy equation, which contains the
sole term that drives the circulation (i.e., radiative heating and
cooling). The quantity $T_{\mathrm{eq}}-T$ on the rightmost side of
\Eq{eq:hpecart3} represents the local difference between the radiative-equilibrium and actual temperature. This difference
varies spatially in value and sign, as it is typically positive on the
dayside and negative on the nightside. Here we seek an expression for
its characteristic magnitude. We note that if $\left| T_{\mathrm{eq}}
- T \right|_{\mathrm{global}}$ is defined as
\begin{equation}
\left| T_{\mathrm{eq}} - T \right|_{\mathrm{global}} \equiv \left| T_{\mathrm{eq}} - T \right|_{\mathrm{day}} + \left| T_{\mathrm{eq}} - T \right|_{\mathrm{night}} \mathrm{,}
\end{equation}
where the differences on the right hand side are characteristic differences for the appropriate hemisphere, one can write
\begin{equation}
\label{eq:deltaT}
\Delta T_{\mathrm{eq}} - \Delta T \sim \left| T_{\mathrm{eq}} - T \right|_{\mathrm{global}} \mathrm{.}
\end{equation}
In Equation (\ref{eq:deltaT}), $\Delta T$ and $\Delta T_{\mathrm{eq}}$ are defined to be the characteristic difference between the dayside and nightside temperature and equilibrium temperature profiles, respectively. \Fig{fig:schematic} shows visually this approximate equality between $\Delta T_{\mathrm{eq}} - \Delta T$ and $\left| T_{\mathrm{eq}} - T \right|_{\mathrm{global}}$.
 \\ \indent With this formalism for characteristic dayside-nightside temperature
differences, we can write an approximate version of \Eq{eq:hpecart3}
as:
\begin{equation}
\label{eq:energfirst}
\frac{\Delta T_{\mathrm{eq}} - \Delta T}{\tau_{\mathrm{rad}}} \sim \mathrm{max} \left[ \frac{\mathcal{U} \Delta T}{\mathcal{L}}\mathrm{,} \frac{\mathcal{W} N^2 H}{R}\right]   \mathrm{.}
\end{equation} 
The quantities $\mathcal{U}$,
$\mathcal{W}$, $\Delta T$, $H$, and $N$ are
implicitly functions of pressure. For the analyses that follow, we use
a value of $\mathcal{L}$ approximately equal to the planetary radius.

What is the relative importance of the two terms on the right-hand side of
(\ref{eq:energfirst})?  Using Equation~(\ref{cont-simple}) and the
definition of Brunt-Vaisala frequency, the second term can be
expressed as $\mathcal{U}/\mathcal{L}$ times $\delta T_{\rm strat}$,
where $\delta T_{\rm strat}$ is the difference between the actual and
adiabatic temperature gradients integrated vertically over a scale
height---or, equivalently, can approximately be thought of as the
change in potential temperature over a scale height.  Thus, the first
term (horizontal entropy advection) dominates over the second term
(vertical entropy advection) only if the day-night temperature (or
potential temperature) difference exceeds the vertical change in potential
temperature over a scale height.  For a highly stratified temperature
profile like those expected in the observable atmospheres of hot
Jupiters, the vertical change in potential temperature over a scale height is a significant fraction of the temperature
itself.\footnote{For a vertically isothermal temperature profile,
  $\delta T_{\rm strat}=gH/c_p = RT/c_p$, and if $R/c_p=2/7$ as
  appropriate for an H$_2$ atmosphere, then $\delta T_{\rm
    strat}\approx 400\rm\,K$ for a typical hot Jupiter with a
  temperature of $1500\rm\,K$.}

Thus, one would expect that vertical entropy advection dominates over
horizontal entropy advection unless the fractional day-night temperature
difference is close to unity.  This is just the weak temperature gradient
regime mentioned in the Introduction.  Considering this WTG balance, we
have\footnote{This balance was first considered for hot Jupiters by
  \citet[][Eq.~22]{showman_2002}.}
\begin{equation}
\label{eq:energhorizapprox}
\frac{\Delta T_{\mathrm{eq}} - \Delta T}{\tau_{\mathrm{rad}}} \sim
\frac{\mathcal{W}N^2 H}{R}.
\end{equation}
Given our full solutions, we will show in Section~\ref{sec:wtg}
that horizontal entropy advection is indeed smaller than vertical
entropy advection in a hemospheric-averaged sense, demonstrating
the validity of (\ref{eq:energhorizapprox}).

\subsubsection{Hydrostatic balance}
 
Hydrostatic balance relates the geopotential to the temperature, and
thus we can use it to relate the day-night temperature difference,
$\Delta T$, to the day-night geopotential difference (alternatively pressure gradient) on isobars.
Hydrostatic balance implies that $\delta \Phi = \int R T
d\mathrm{ln}p$, where here $\delta \Phi$ is a {\it vertical}
geopotential difference. Then, consider evaluating $\delta \Phi$ on
the dayside and nightside, for two vertical air columns sharing the same value of $\Phi$ at their base, at points separated by a horizontal
distance $\mathcal{L}$. Given that these two points have the same
$\Phi$ at the bottom isobar, we can difference $\delta \Phi$ at these
locations to solve for the geopotential change from dayside to
nightside. Doing so, we find the horizontal geopotential
difference\footnote{Note that this is essentially the hypsometric
  equation (e.g. \citealp{Holton:2013} pp. 19 or
  \citealp{Wallace:2004} pp. 69-72).}:
\begin{equation}
\label{eq:pressgrad}
\Delta \Phi \approx R \int_p^{p_{\mathrm{bot}}} \Delta T d\mathrm{ln}p' \mathrm{.}
\end{equation} 
In \Eq{eq:pressgrad}, both $\Delta\Phi$ and $\Delta T$ are functions of pressure. Here we only have to integrate to the pressure level at which our
prescribed equilibrium dayside-nightside temperature difference
$\Delta T_{\mathrm{eq}}$ goes to zero, which is labeled
$p_{\mathrm{bot}}$.

Given that differences in scalar quantities from dayside to nightside are a function of pressure in our model, we can use our Newtonian cooling parameterizations as a guide to the form this pressure-dependence will take. Hence, we take a form of $\Delta T$ similar to that of $\Delta T_{\mathrm{eq}}$, focusing only on the region with pressure dependence:
\begin{equation}
\Delta T = \Delta T_{\mathrm{top}} \frac{\mathrm{ln}(p/p_{\mathrm{bot}})}{\mathrm{ln}(p_{\mathrm{eq,top}}/p_{\mathrm{bot}})} \mathrm{.}
\end{equation}
Now, integrating Equation (\ref{eq:pressgrad}) and dropping a factor of two, we find 
\begin{equation}
\label{eq:deltageopot}
\Delta \Phi \approx R \Delta T \mathrm{ln}\left(\frac{p_{\mathrm{bot}}}{p}\right) \mathrm{.}
\end{equation}

\subsubsection{Momentum equation}
\label{sec:momequationsetup}
Next we analyze the approximate horizontal momentum equation, where the pressure gradient force driving the circulation can be balanced by horizontal advection (in regions where the Rossby number $\mathrm{Ro} \equiv \mathcal{U}/f\mathcal{L} \gg 1)$, vertical advection, the Coriolis force ($\mathrm{Ro} \ll 1$), or drag:
\begin{equation}
\label{eq:horizmom}
\nabla \Phi \sim \mathrm{max} \left[\frac{\mathcal{U}^2}{\mathcal{L}}\mathrm{,} \frac{\mathcal{U}\mathcal{W}}{H} \mathrm{,} f\mathcal{U} \mathrm{,} \frac{\mathcal{U}}{\tau_{\mathrm{drag}}} \right]\mathrm{.}
\end{equation}

Note that, given our continuity equation (\ref{cont-simple}), the horizontal
and vertical momentum advection terms $\mathcal{U}^2/\mathcal{L}$ and 
$\mathcal{U W}/H$ are identical.  As a result, we expect that the final
solutions for $\Delta T(p)$, $\mathcal{U}(p)$, and $\mathcal{W}(p)$ should be
the same if horizontal momentum advection balances the pressure gradient force as for the
case when vertical momentum advection balances it.

Expressing $\nabla\Phi$ as $\Delta \Phi/\mathcal{L}$, and
  relating $\Delta\Phi$ to $\Delta T$ using (\ref{eq:deltageopot}), we
  evaluate Equation~(\ref{eq:horizmom}) for every possible dominant
  term on the right hand side, now with explicit pressure-dependence.
  This leads to approximate horizontal velocities that scale as

\begin{equation}
\label{eq:ucalc}
\everymath=\expandafter{\displaystyle}
\mathcal{U}(p) \sim \begin{cases} \frac{R \Delta T(p)\Delta \mathrm{ln}p \  \tau_{\mathrm{drag}}(p)}{\mathcal{L}}  \hspace{0.8cm} \mathrm{Drag} \\ 
\frac{R \Delta T(p)\Delta \mathrm{ln}p}{\mathcal{L} f} \hspace{2.cm} \mathrm{Coriolis} \\ 
\sqrt{R \Delta T(p)\Delta \mathrm{ln}p} \hspace{1.8cm} \mathrm{Advection}\mathrm{,}
\end{cases}
\end{equation}
where we define $\Delta \ln p =\ln(p_{\rm bot}/p)$ as the
  difference in log pressure from the deep pressure $p_{\rm bot} = 10 \ \mathrm{bars}$, where the day-night forcing goes to zero, to
  some lower pressure of interest. As foreshadowed above, the same solution,
given by the final expression in (\ref{eq:ucalc}), 
is attained when either horizontal or vertical momentum advection
balances the pressure gradient force.  The expressions for $\mathcal{U}$ in
\Eq{eq:ucalc} for advection and Coriolis force balancing
pressure gradient match those of \cite{Showman_2009} (see their
Equations 48 and 49). Hence, we recapture within our idealized model
the previous expectations for characteristic horizontal wind speeds on
hot Jupiters based on simple force balance. Note that these are
expressions for characteristic horizontal wind speed, which represents
average day-night flow and hence is not the zonal-mean zonal wind
associated with superrotation.

Invoking the relationship between $\mathcal{U}$ and $\mathcal{W}$ from Equation~(\ref{cont-simple}),
we can now write expressions for the vertical wind speed as a function of $\Delta T$:
\begin{equation}
\label{eq:wcalc}
\everymath=\expandafter{\displaystyle}
\mathcal{W}(p) \sim \begin{cases} \frac{R \Delta T(p) H(p) \Delta \mathrm{ln}p \  \tau_{\mathrm{drag}}(p)}{\mathcal{L}^2}  \hspace{0.8cm} \mathrm{Drag} \\ 
\frac{R \Delta T(p) H(p) \Delta\mathrm{ln}p}{\mathcal{L}^2 f} \hspace{2.cm} \mathrm{Coriolis} \\ 

{H(p)\over \mathcal{L}}\sqrt{R \Delta T(p)\Delta \mathrm{ln}p} \hspace{1.7cm} \mathrm{Advection}\mathrm{.} 
\end{cases}
\end{equation}

\subsection{Full Solution for fractional dayside-nightside temperature differences and wind speeds}
\label{sec:finalresult}

 To obtain our final solution, we simply combine
  Equation~(\ref{eq:energhorizapprox}) and (\ref{eq:wcalc}) to yield
  expressions for $\Delta T(p)$ and $\mathcal{W}(p)$ as a function of
  control parameters:
\begin{equation}
\label{eq:tempfinal}
\everymath=\expandafter{\displaystyle}
\frac{\Delta T(p)}{\Delta T_{\mathrm{eq}}(p)} \sim 	\begin{cases} \left(1 +  \frac{\tau_{\mathrm{rad}}(p) \tau_{\mathrm{drag}}(p)}{\tau_{\mathrm{wave}}^2(p)}\Delta \mathrm{ln}p\right)^{-1}	\hspace{.0cm} \mathrm{Drag}\\ \left( 1 +  \frac{\tau_{\mathrm{rad}}(p)}{f \tau_{\mathrm{wave}}^2(p)}\Delta \mathrm{ln}p\right)^{-1}\hspace{0.7cm} \mathrm{Coriolis}\\ \frac{\sqrt{\gamma(p) + 4 \Delta T_{\mathrm{eq}}(p)} - \sqrt{\gamma(p)}}{\sqrt{\gamma(p) + 4 \Delta T_{\mathrm{eq}}(p)} + \sqrt{\gamma(p)}} \hspace{.3cm} \mathrm{Advection}\mathrm{,} \end{cases} 
\end{equation}
and
\begin{equation}
\label{eq:Wfinal}
\begin{aligned}
\everymath=\expandafter{\displaystyle}
\mathcal{W} \sim 	\begin{cases} \frac{R H(p)}{\mathcal{L}^2}\frac{\tau_{\mathrm{wave}}^2(p)}{\tau_{\mathrm{rad}}(p)} \Delta T_{\mathrm{eq}}(p) \\ \times \left[1 - \left(1 +  \frac{\tau_{\mathrm{rad}}(p) \tau_{\mathrm{drag}}(p)}{\tau_{\mathrm{wave}}^2(p)}\Delta \mathrm{ln}p\right)^{-1}\right]	\hspace{.0cm} \mathrm{Drag}\\ \\ \frac{R H(p)}{\mathcal{L}^2}\frac{\tau_{\mathrm{wave}}^2(p)}{\tau_{\mathrm{rad}}(p)} \Delta T_{\mathrm{eq}}(p) \\ \times \left[1 - \left( 1 +  \frac{\tau_{\mathrm{rad}}(p)}{f \tau_{\mathrm{wave}}^2(p)}\Delta \mathrm{ln}p\right)^{-1}\right]\hspace{0.7cm} \mathrm{Coriolis}\\ \\ \frac{R H(p)}{\mathcal{L}^2}\frac{\tau_{\mathrm{wave}}^2(p)}{\tau_{\mathrm{rad}}(p)} \Delta T_{\mathrm{eq}}(p) \\ \times \left[1 - \frac{\sqrt{\gamma(p) + 4 \Delta T_{\mathrm{eq}}(p)} - \sqrt{\gamma(p)}}{\sqrt{\gamma(p) + 4 \Delta T_{\mathrm{eq}}(p)} + \sqrt{\gamma(p)}}\right]\hspace{.3cm} \mathrm{Advection}\mathrm{.} \end{cases} 
\end{aligned}
\end{equation}
The solution for $\mathcal{U}(p)$ is simply that for $\mathcal{W}(p)$
times $\mathcal{L}/H$.

In these solutions, we have defined $\gamma(p) = \tau^2_{\mathrm{rad}}(p)
\mathcal{L}^2\Delta \mathrm{ln}p/(\tau^4_{\mathrm{wave}}(p)R)$.
Moreover, we have substituted a Kelvin wave propagation timescale
$\tau_{\mathrm{wave}}(p) = \mathcal{L}/(\mathcal{N}(p)H(p))$, as in
\cite{showman_2013_doppler}. This wave propagation timescale is
derived by taking the long vertical wavelength limit of the Kelvin
wave dispersion relationship, giving the fastest phase and group
propagation speeds $c = 2NH$. Hence, the wave propagation time across
a hemisphere is approximately $\mathcal{L}/(\mathcal{N}H)$.

Our drag and Coriolis-dominated equations are the same expressions as
in \cite{Perez-Becker:2013fv}, but the advection-dominated cases
differ. Note that as in \cite{Perez-Becker:2013fv}, the
dayside-nightside temperature (thickness in their case) differences
for the drag and Coriolis-dominated regimes are equal if
$\tau_{\mathrm{drag}} \sim 1/f$.  The solutions for $\Delta T/\Delta T_{\rm eq}$
in the drag and Coriolis cases are independent of the forcing amplitude
$\Delta T_{\rm eq}$, whereas the solution in the advection case depends
on $\Delta T_{\rm eq}$, as one would expect given the nonlinear nature of
the momentum balance.



With these final solutions for the dayside-nightside temperature difference and characteristic wind speeds, we can compare our theory to the GCM results in \Sec{sec:results}. However, we first must diagnose which solution is appropriate for any given combination of parameters.

\subsection{Momentum equation regime} 
\label{sec:momeqregime}
To determine which regime is appropriate in our solutions from \Sec{sec:finalresult} for given choices of control parameters ($\tau_{\mathrm{rad,top}}$, $\tau_{\mathrm{drag}}$, and $\Delta T_{\rm eq,top}$), we perform a comparison of the magnitudes of the drag, Coriolis, and momentum advection terms. We do this by comparing the relative amplitues of their characteristic timescales: $\tau_{\mathrm{drag}}$, $1/f$, and $\tau_{\mathrm{adv}}$. Using Equations (\ref{cont-simple}) and (\ref{eq:Wfinal}), we can write $\tau_{\mathrm{adv}}$ as
\begin{equation}
\label{eq:tauadv}
\begin{split}
\tau_{\mathrm{adv}}(p) = & \frac{\mathcal{L}^2 \tau_{\mathrm{rad}}(p)}{R \tau^2_{\mathrm{wave}}(p) \Delta T_{\mathrm{eq}}(p)} \\ & \times \left[1 - \frac{\sqrt{\gamma(p) + 4 \Delta T_{\mathrm{eq}}(p)} - \sqrt{\gamma(p)}}{\sqrt{\gamma(p) + 4 \Delta T_{\mathrm{eq}}(p)} + \sqrt{\gamma(p)}}\right]^{-1} \mathrm{.}
\end{split}
\end{equation}

Now, we can write the condition that determines which regime the solution is in. If $\tau_{\mathrm{drag}} < \mathrm{min}(1/f,\tau_{\mathrm{adv}})$, the solution is in the drag-dominated regime. If $1/f < \mathrm{min}(\tau_{\mathrm{drag}},\tau_{\mathrm{adv}})$, it is in the Coriolis regime. If $\tau_{\mathrm{adv}} < \mathrm{min}(1/f,\tau_{\mathrm{drag}})$, it is in the advection-dominated regime.

To evaluate these conditions, we proceed as follows.  One can
  start by evaluating the Rossby number $\mathrm{Ro}=\mathcal{U}/f\mathcal{L}$
  using Eqs.~(\ref{cont-simple}) and (\ref{eq:Wfinal}) for given
  external parameters, and this determines the relative magnitudes of
  Coriolis and advective forces\footnote{Alternatively, one can calculate the Rossby number from the advective timescale, as we take $f$ as an external parameter and $\mathcal{U}/\mathcal{L} = \tau_{\mathrm{adv}}^{-1}$. Hence, one can calculate the Rossby number (as a function of latitude) as $\mathrm{Ro}(\phi) = \left[f(\phi)\tau_{\mathrm{adv}}\right]^{-1}$, where $\tau_{\mathrm{adv}}$ is given by \Eq{eq:tauadv}.}.  First, we consider the regime where
$\mathrm{Ro} < 1$, which corresponds to the regions where the Coriolis
force has a magnitude larger than that of advection. This is
appropriate in the mid-to-high latitudes of a planet, with the
latitudinal extent of this regime increasing with increasing rotation
rate. Though the Coriolis force dominates over advection, if
$\tau_{\mathrm{drag}}$ is very short, drag can have a larger magnitude
than the Coriolis force. Hence, we consider a three-term force balance
in this regime, with drag balancing pressure gradient if
$\tau_{\mathrm{drag}} < 1/f$ and Coriolis balancing the pressure gradient if $1/f
< \tau_{\mathrm{drag}}$. Given this, we can write a solution for the
fractional day-night temperature difference $A(p) \sim \Delta
T(p)/\Delta T_{\mathrm{eq}}(p)$ (labeled such for future comparison with $A_{\mathrm{obs}}$ in \Sec{sec:results}) in the $\mathrm{Ro} < 1$ regime:
\begin{equation}
\label{overallAtheory}
\everymath=\expandafter{\displaystyle}
A(p) \sim \begin{cases} \left(1 +  \frac{\tau_{\mathrm{rad}}(p) \tau_{\mathrm{drag}}(p)}{\tau_{\mathrm{wave}}^2(p)}\Delta \mathrm{ln}p\right)^{-1} \hspace{0.05cm} \tau_{\mathrm{drag}} < f^{-1} \\ \left( 1 +  \frac{\tau_{\mathrm{rad}}(p)}{f \tau_{\mathrm{wave}}^2(p)}\Delta \mathrm{ln}p\right)^{-1} \hspace{0.8cm} \tau_{\mathrm{drag}} > f^{-1}\mathrm{.} \end{cases}
\end{equation}
\indent In the regime where $\mathrm{Ro} > 1$, the advection term has a greater magnitude than the Coriolis force. This is relevant to the equatorial regions of any given planet, where $f \rightarrow 0$ due to the latitudinal dependence of the Coriolis force. Additionally, if advection is very strong (or rotation very slow), this regime could extend to the mid-to-high latitudes. The three-term force balance in regions where $\mathrm{Ro} > 1$ is hence between drag, advection, and the day-night pressure gradient force. Using the same reasoning as above, we can write the fractional day-night temperature difference at the equator $A_{\mathrm{eq}}(p)$:
\begin{equation}
\label{eqAtheory}
\everymath=\expandafter{\displaystyle}
A_{\mathrm{eq}}(p) \sim \begin{cases} \\ \left(1 +  \frac{\tau_{\mathrm{rad}}(p) \tau_{\mathrm{drag}}(p)}{\tau_{\mathrm{wave}}^2(p)}\Delta \mathrm{ln}p\right)^{-1} \hspace{-0.3cm}\tau_{\mathrm{drag}}(p) < \tau_{\mathrm{adv}}(p) \\ \\ \frac{\sqrt{\gamma(p) + 4 \Delta T_{\mathrm{eq}}(p)} - \sqrt{\gamma(p)}}{\sqrt{\gamma(p) + 4 \Delta T_{\mathrm{eq}}(p)} + \sqrt{\gamma(p)}} \hspace{0cm} \tau_{\mathrm{drag}}(p) > \tau_{\mathrm{adv}}(p) \mathrm{,}  \end{cases}
\end{equation}
where $\tau_{\rm adv}$ is evaluated from Eq.~(\ref{eq:tauadv}).

\subsection{Transition from low to high day-night temperature differences} 
\label{sec:transdT}

From the expressions above for the fractional day-night temperature
difference, it is clear that a comparison between wave propagation
timescales and other relevant timescales (radiative, drag, Coriolis)
controls the amplitude of the day-night temperature difference. 
Here, we use our theory to obtain timescale comparisons for the transition
from small to large day-night temperature difference (relative to that in
radiative equilibrium).

Using the expression for overall $A$ from \Eq{overallAtheory}, the
transition between low fractional dayside-nightside temperature
differences ($A \rightarrow 0$) and high fractional dayside-nightside
temperature differences ($A \rightarrow 1$) occurs when\footnote{Note
  that the two expressions in \Eq{eq:transitionoverall} are equivalent
  to a pressure-dependent versions of Equations (24) in
  \cite{Perez-Becker:2013fv}}:
\begin{equation}
\label{eq:transitionoverall}
\everymath=\expandafter{\displaystyle}
\tau_{\mathrm{wave}}(p) \sim \begin{cases} \sqrt{\tau_{\mathrm{drag}}(p)\tau_{\mathrm{rad}}(p)\Delta \mathrm{ln}p}  \hspace{0.5cm}  \tau_{\mathrm{drag}}(p) < f^{-1} \\ \sqrt{f^{-1} \tau_{\mathrm{rad}}(p)\Delta \mathrm{ln}p} \hspace{1.05cm} \tau_{\mathrm{drag}}(p) > f^{-1}\mathrm{.} \end{cases}
\end{equation}
This timescale comparison is valid in regions where $\mathrm{Ro} < 1$, which occurs at nearly all latitudes (except those bordering the equator, where $A$ is given by Equation \ref{eqAtheory}). When $\tau_{\mathrm{wave}}$ is smaller than the expression on the righthand side of \Eq{eq:transitionoverall}, the day-night temperature differences are small, while if it is larger the day-night temperature differences are necessarily large. It is notable that the transition between low and high day-night temperature difference always involves a comparison between wave and radiative timescales, no matter what regime the system is in. Meanwhile, drag is only important if $\tau_{\mathrm{drag}} < \mathrm{min}(1/f,\tau_{\mathrm{adv}})$, which is why drag only plays a role in determining the day-night temperature differences if $\tau_{\mathrm{drag}} < 1/f \sim 10^5 \ \mathrm{sec}$ for our choice of $\Omega$. 

We can motivate the timescale comparison in \Eq{eq:transitionoverall} by considering the time it takes for adiabatic vertical motions to smooth out a day-night temperature difference equal to $\Delta T$. The contribution of vertical motion to the time variation of temperature is $\mathcal{W}N^2H/R$ (see Equation 23), which implies that we can express the timescale $\tau_{\mathrm{vert}}$ for day-night temperature differences to be erased by vertical entropy advection as:
\begin{equation}
\label{eq:tauvert}
\tau_{\mathrm{vert}} \sim \frac{\Delta T R}{\mathcal{W}N^2H} \mathrm{.}
\end{equation}
For concreteness, consider the drag-dominated regime where $\tau_{\mathrm{drag}} < \min(1/f,\tau_{\mathrm{adv}})$. Substituting our solutions for $\Delta T$ and $\mathcal{W}$ from Equations (\ref{eq:tempfinal}) and (\ref{eq:Wfinal}), respectively, we can write:
\begin{equation}
\label{eq:tauvert}
\tau_{\mathrm{vert}} \sim \frac{\tau^2_{\mathrm{wave}}}{\tau_{\mathrm{drag}}\Delta\mathrm{ln}p} \mathrm{.}
\end{equation}
If $\tau_{\mathrm{rad}}$ is on the order of $\tau_{\mathrm{vert}}$, the approximate equality
\begin{equation}
\label{eq:tauvert}
\tau_{\mathrm{rad}} \sim \frac{\tau^2_{\mathrm{wave}}}{\tau_{\mathrm{drag}}\Delta\mathrm{ln}p}
\end{equation}
holds, which is precisely the same timescale comparison as in \Eq{eq:transitionoverall} for the strong-drag regime. Performing the same exercise for the regime where $1/f < \mathrm{min}(\tau_{\mathrm{drag}},\tau_{\mathrm{adv}})$, 
\begin{equation}
\label{eq:tauvert}
\tau_{\mathrm{vert}} \sim \tau_{\mathrm{rad}} \sim \frac{f \tau^2_{\mathrm{wave}}}{\Delta\mathrm{ln}p} \mathrm{.}
\end{equation}
This is precisely the same timescale comparison as in \Eq{eq:transitionoverall} for the weak-drag regime. Note that if $\tau_{\mathrm{vert}} \ll \tau_{\mathrm{rad}}$, the solution approaches the limit of small $A$, with $\tau_{\mathrm{wave}} \ll \mathrm{min}\left[\sqrt{f^{-1}\tau_{\mathrm{rad}}\Delta\mathrm{ln}p},\sqrt{\tau_{\mathrm{drag}}\tau_{\mathrm{rad}}\Delta\mathrm{ln}p}\right]$. In the other limit of $\tau_{\mathrm{vert}} \gg \tau_{\mathrm{rad}}$, the solution approaches that in radiative equilibrium, with $A \rightarrow 1$. This link between $\tau_{\mathrm{vert}}$ and $\tau_{\mathrm{wave}}$ is a natural consequence of the WTG limit, where vertical motion that moves isentropes vertically (parameterized here by $\tau_{\mathrm{vert}}$) is forced by horizontal convergence/divergence caused by propagating waves (parameterized by $\tau_{\mathrm{wave}}$).  
\section{Test of the Theory with 3D Simulations}
\label{sec:results}
\subsection{Computing fractional dayside-nightside temperature differences from numerical results}
Though we have expressions for the dayside-nightside temperature difference relative to the equilibrium dayside-nightside temperature difference, we must relate this to a metric we can compute using our numerical solutions. To compute this metric for the overall dayside-nightside temperature contrast, we follow a method similar to \cite{Perez-Becker:2013fv}. First, we evaluate the longitudinal RMS variations in temperature, to find $A(\phi,p)$, the overall fractional dayside-nightside temperature difference as a function of latitude and pressure:
\begin{equation}
\label{eq:Adef}
A(\phi,p) = \left \{\frac{\int_{0}^{2\pi}	 \left[T(\lambda,\phi,p) - \bar{T}(\phi,p)\right]^2 d\lambda}{\int_{0}^{2\pi}	 \left[T_{\mathrm{eq}}(\lambda,\phi,p) - \bar{T}(\phi,p)\right]^2 d\lambda}	\right \}^{1/2} \mathrm{.}
\end{equation}
Here $\bar{T}$ is the zonal-mean temperature at a given latitude and pressure. Then, we average $A(\phi,p)$ with latitude over a $120^{\circ}$ band, centered around the equator\footnote{ We do not include the poles in this average, as the dayside-nightside forcing is strongest at low latitudes in the Newtonian cooling scheme used due to the $\mathrm{cos}(\phi)$ dependence of radiative-equilibrium temperature. Additionally, their larger projected area implies that low latitudes dominate the emitted flux seen in phase curves.} to find $A$, a numerical representation of $\Delta T/\Delta T_{\mathrm{eq}}$:
\begin{equation}
A(p) = \frac{3}{2\pi} \int_{-\pi/3}^{\pi/3} A(\phi,p) d\phi {\mathrm{.}}
\end{equation}
After this averaging, we choose a nominal pressure level, $\sim 80$ mbars, to calculate near-photospheric dayside-nightside temperature differences. This $A$ is relevant to the $\mathrm{Ro} <1$ limit, where \Eq{overallAtheory} gives our theoretical prediction for $A$. To compare with the predictions for day-night temperature differences at the equator, given by \Eq{eqAtheory}, we simply evaluate $A(\phi,p)$ from \Eq{eq:Adef} at the equator itself, where $\phi = 0^{\circ}$. We also take the pressure-dependence of $A$ into account, which informs us about how dayside-nightside temperature differences vary with depth. Now we can directly compare our theoretical expressions for $A(p)$ with numerical solutions.
\begin{figure*}[p]
	\centering
	\includegraphics[width=.70\textwidth]{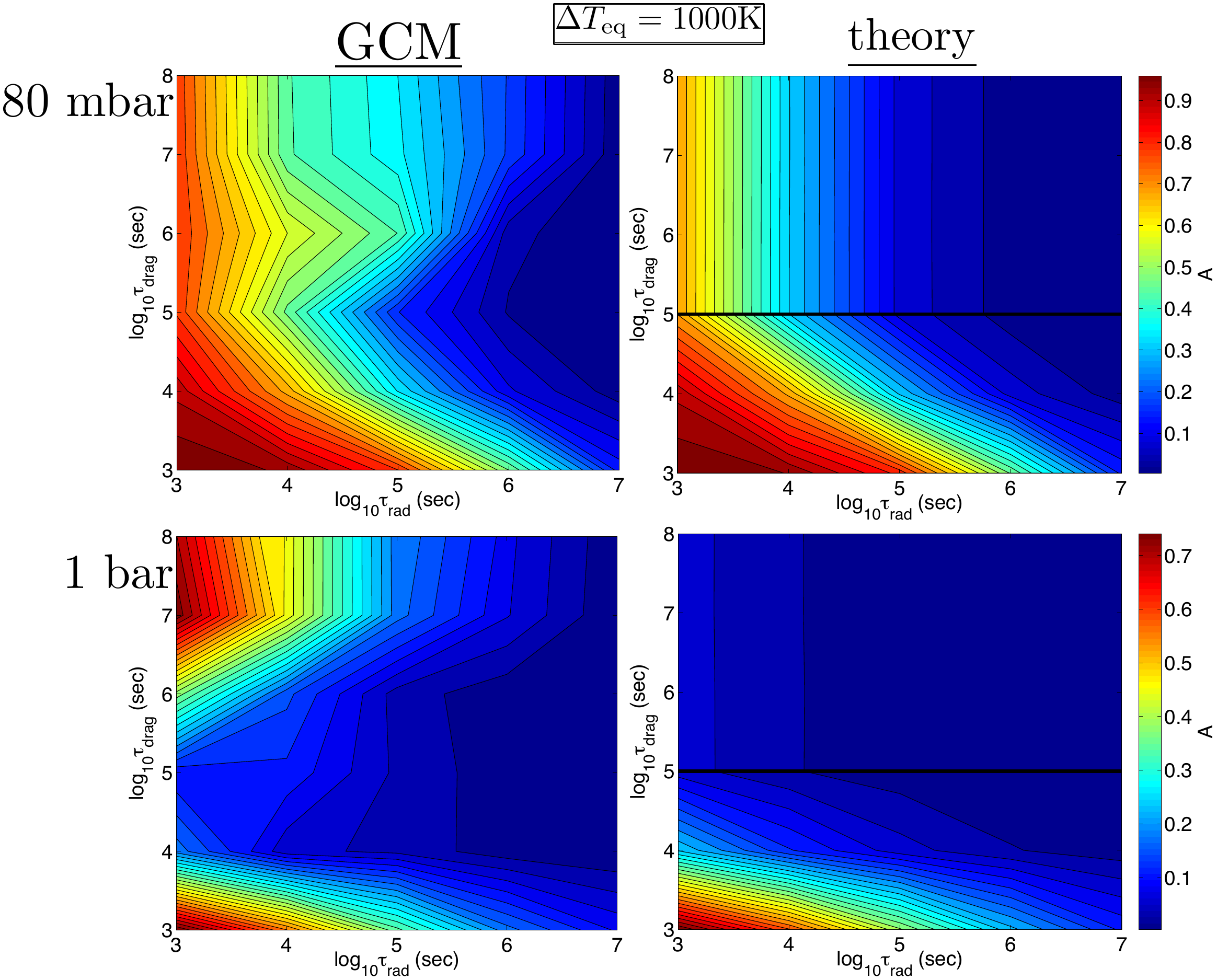}
	\caption{Overall fractional dayside-nightside temperature temperature difference, $A$, plotted as a function of $\tau_{\mathrm{rad,top}}$ (abscissa) and $\tau_{\mathrm{drag}}$ (ordinate) for both GCM results (left) and theoretical predictions (right), and two different pressure levels: 80 mbars (top row), 1 bar (bottom row). The model and theory plots for a given pressure level have the same color scale. The theoretical predictions are based on the competition between pressure-gradient forces and either drag or Coriolis forces, see Equation (\ref{overallAtheory}). The black line on the righthand plots demarcates the regions where drag or Coriolis forces dominate: above the black line, Coriolis force is balanced with pressure gradient, and below drag is the dominant balancing term. The colormap is scaled such that red corresponds to high temperature differences and blue to small temperature differences.}
	\label{fig:Acompare_high}
\end{figure*}
\begin{figure*}[p]
	\centering
	\includegraphics[width=.70\textwidth]{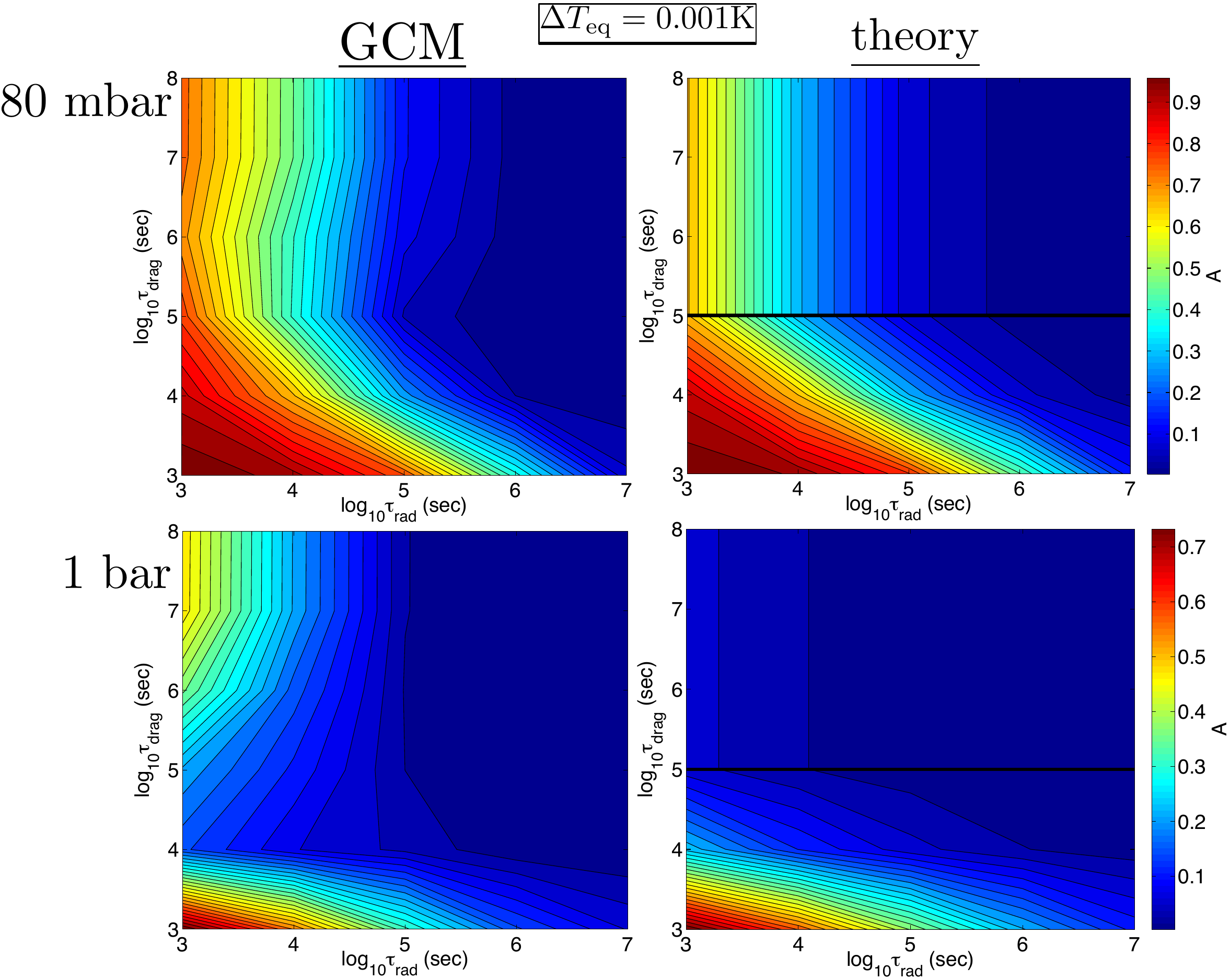}
	\caption{As in Figure \ref{fig:Acompare_high}, but for the case with $\Delta T_{\mathrm{eq,top}} = 0.001$ Kelvin.}
	\label{fig:Acompare_low}
\end{figure*}
\begin{figure*}[p]
	\centering
	\includegraphics[width=.70\textwidth]{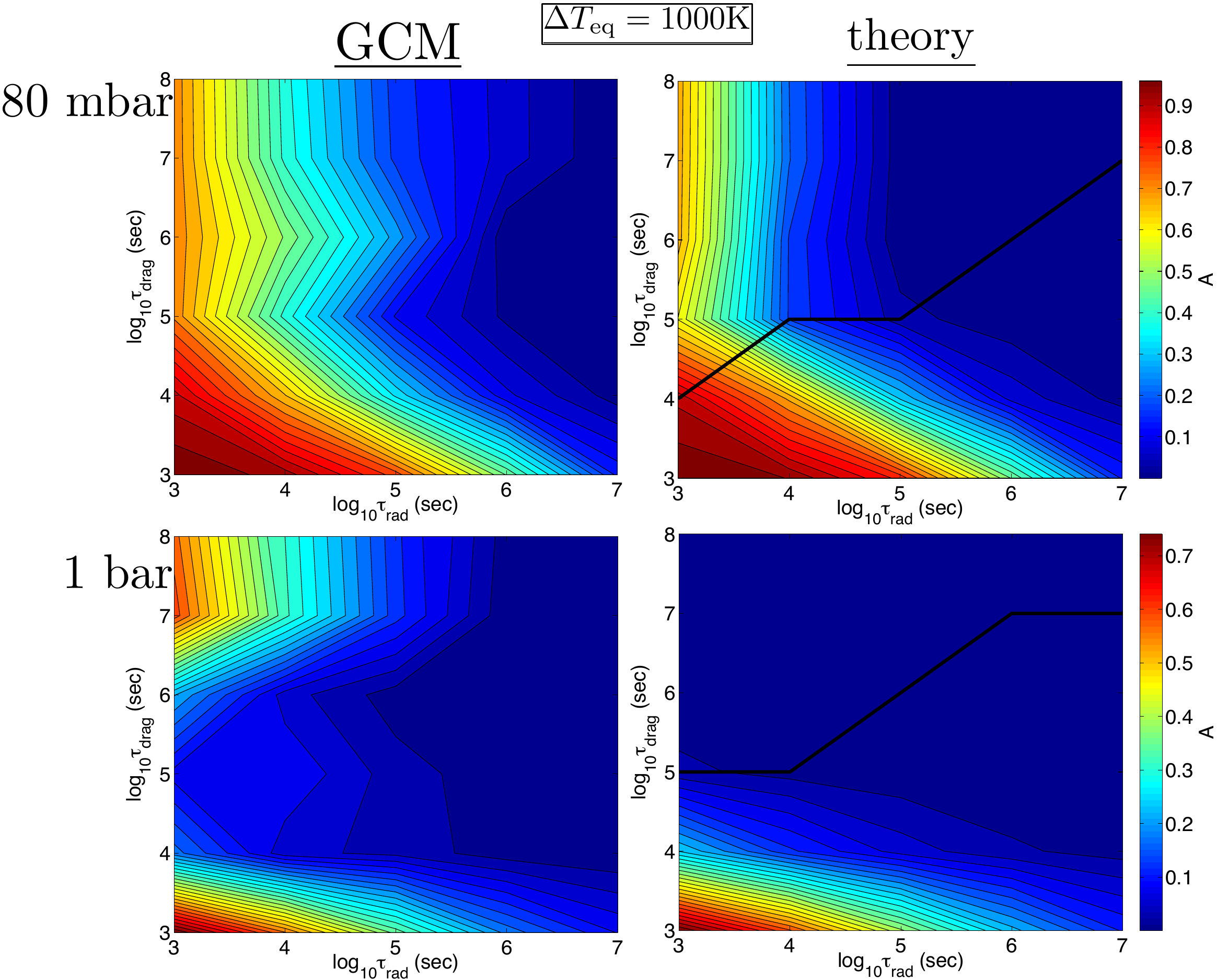}
	\caption{As in Figure \ref{fig:Acompare_high}, but with $A$ only evaluated at the planetary equator. In this figure, the theoretical predictions are based on the competition between drag and advection, see Equation (\ref{eqAtheory}). The black line demarcates this transition: above the black line, advection is the dominant balance with the pressure gradient force, while below the black line drag is the controlling term.}
	\label{fig:Acompare_eqhigh}
\end{figure*}
\begin{figure*}[p]
	\centering
	\includegraphics[width=.70\textwidth]{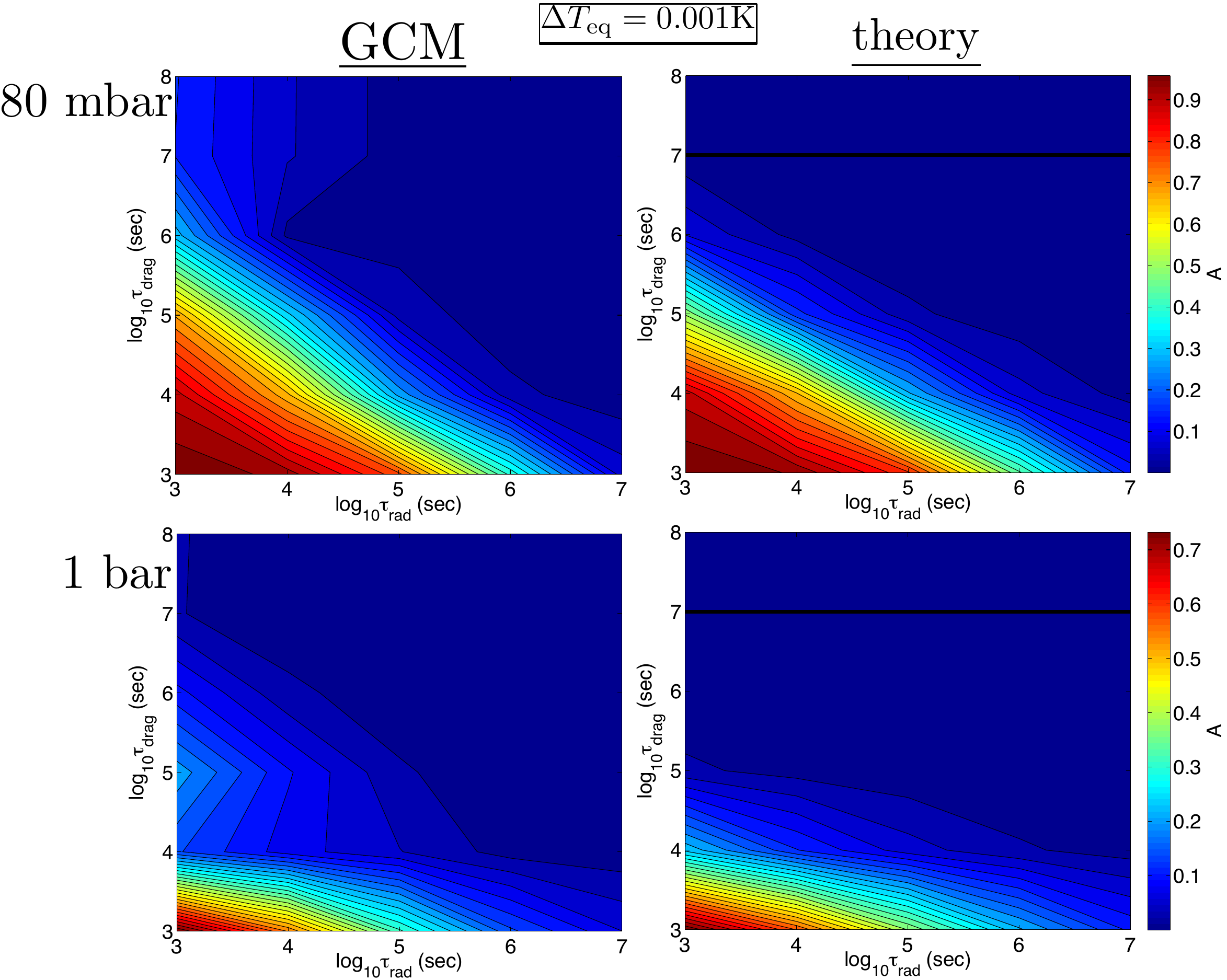}
	\caption{As in Figure \ref{fig:Acompare_eqhigh}, but for the case with $\Delta T_{\mathrm{eq,top}} = 0.001$ Kelvin.}
	\label{fig:Acompare_eqlow}
\end{figure*}
 \begin{figure*}
	\centering
	\includegraphics[height=.7\textheight]{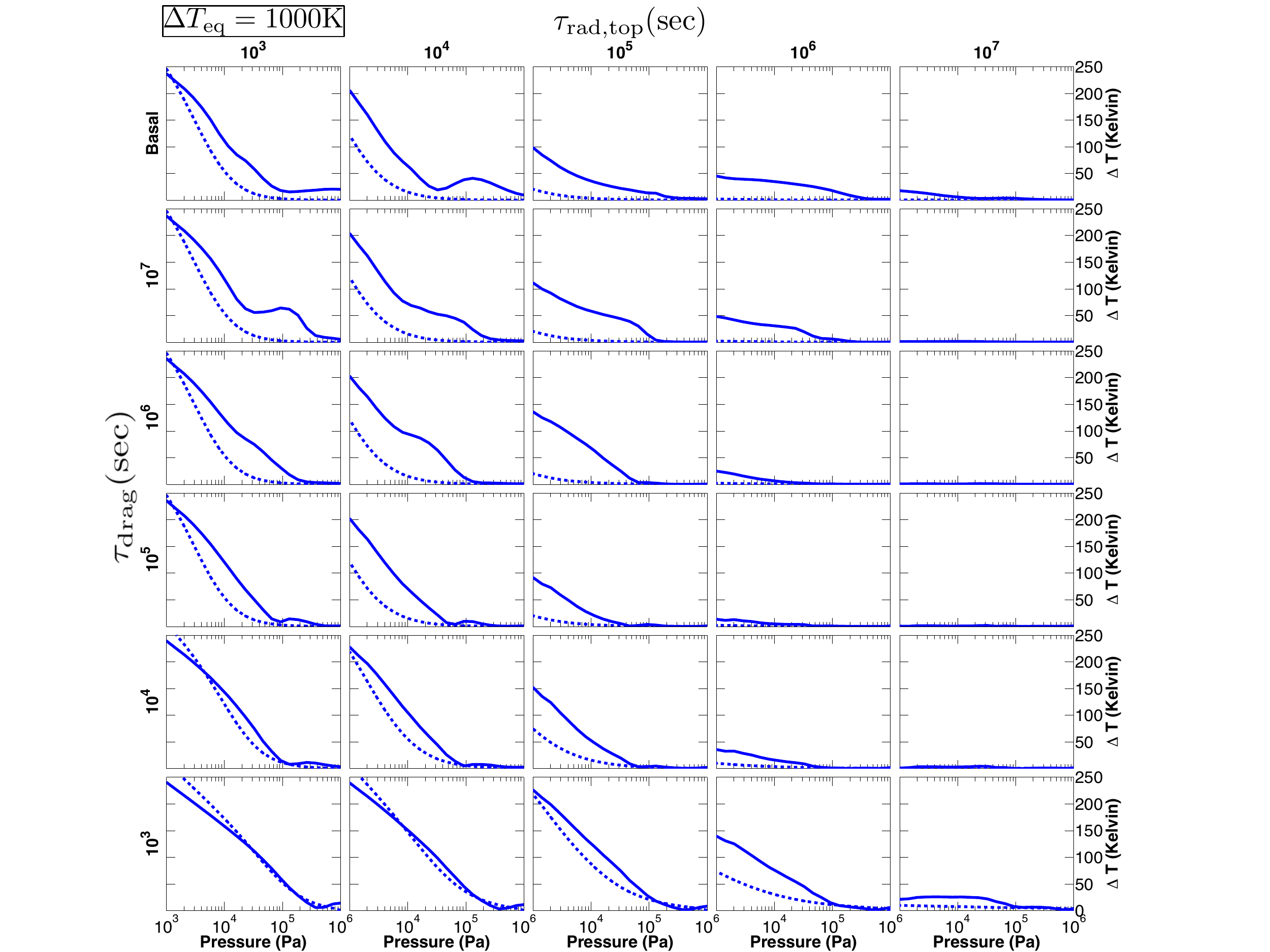}
	\caption{Comparison of theoretical prediction (dashed lines) and GCM results (solid lines) for characteristic overall  dayside-nightside temperature difference $\Delta T = T_{\mathrm{RMS,day}} - T_{\mathrm{RMS,night}}$ as a function of pressure for the case where $\Delta T_{\mathrm{eq,top}} = 1000 \hspace{2pt} \mathrm{Kelvin}$. Since our theory predicts that the characteristic Rossby number is $\lesssim 1$ over most of the parameter space, the theory curves utilize the three-way force balance between Coriolis, drag, and pressure-gradient forces in the momentum equation. ``Basal'' here means $\tau_{\mathrm{drag}} = \infty$, and only the Held-Suarez drag applies. Note that the theory curves in a given column when $\tau_{\mathrm{drag}} \gtrsim 10^5 \hspace{2pt} \mathrm{sec}$ are similar, which is a statement that the wave propagation timescales are approximately equivalent. There is good agreement between the numerical solutions and theoretical predictions for $\Delta T$ over much of the parameter space. The dayside-nightside temperature difference decreases with increasing pressure in both the theoretical predictions and numerical simulations.}
	\label{fig:apress}
\end{figure*}
\subsection{Comparison of theory to numerical results}
\label{sec:theorycompare}
\indent We now use our expectations for $A$ from \Sec{sec:momeqregime} to directly compare model results and theoretical expectations. First, we compare the model overall $A$ from the theoretical predictions in Equation (\ref{overallAtheory}) for the $\Delta T_{\mathrm{eq,top}} = 1000 \hspace{2pt} \mathrm{Kelvin}$  case at two pressure levels, 80 mbars and 1 bar, shown in \Fig{fig:Acompare_high}. We show the same comparison for the $\Delta T_{\mathrm{eq,top}} = 0.001 \hspace{2pt} \mathrm{Kelvin}$ case in \Fig{fig:Acompare_low}. These are predictions that use a momentum balance between either the Coriolis force and pressure gradient or drag and pressure gradient. In this and all the comparisons that follow, we calculate $\tau_{\mathrm{wave}}$ from the local scale height and vertically-isothermal limit of the Brunt-Vaisala frequency. The black line on the theory plots demarcates the transition between these balances. Above the black line, the Coriolis force balances pressure gradient, while below the black line drag balances the pressure-gradient force. \\
\indent From Figures \ref{fig:Acompare_high} and \ref{fig:Acompare_low}, the similarity between the GCM results and theory show that our analysis quantitatively reproduces the overall dayside-nightside temperature differences well. In both the numerical results and theory, fractional dayside-nightside temperature differences decrease as $\tau_{\mathrm{rad}}$ increases. Additionally, if drag is stronger than the Coriolis force, the dayside-nightside temperature differences are larger. That the transition between high and low dayside-nightside temperature differences depends on both $\tau_{\mathrm{drag}}$ and $\tau_{\mathrm{rad}}$ in the region where drag is stronger than the Coriolis force can be understood from \Eq{eq:transitionoverall}. Here, $A(p)$ is inversely correlated with both $\tau_{\mathrm{drag}}$ and $\tau_{\mathrm{rad}}$, while above the black line (where the Coriolis force is stronger than drag) $A(p)$ only depends on $\tau_{\mathrm{rad}}$. \\
The main discrepancy between the theoretical prediction and GCM results for overall dayside-nightside temperature differences occurs at high pressures ($\sim 1 \hspace{2pt} \mathrm{bar}$) when radiative forcing is strong. This can be seen in the upper-left hand quadrant of the 1 bar comparison in Figures \ref{fig:Acompare_high} and \ref{fig:Acompare_low}. This is due to the circulation (Gill) pattern, which allows for a significant horizontal temperature contrast in these cases. In this region of $\tau_{\mathrm{rad}},\tau_{\mathrm{drag}}$ parameter space, winds do not necessarily blow straight from day to night across isotherms as assumed in our theory, but circle around hot/cold regions in large-scale vortices. This can be seen at lower pressures in Figures \ref{fig:nltempgrid} and \ref{fig:lintempgrid}, and persists to $\sim 1$ bar pressure. Similar circulation patterns have been seen previously, for example in the simulations of \cite{Showman:2008} using Newtonian cooling and by \cite{Showmanetal_2009} when including non-grey radiative transfer. Hence, at pressures of $\sim 1 \hspace{2pt} \mathrm{bar}$, our theory (Equation \ref{eq:tempfinal}) does not predict the day-night temperature difference well in the weak-drag, strong-radiation corner of the parameter space. However, at even greater pressures this pattern does not exist and $A(p)$ again becomes a proper metric for the dayside-nightside temperature contrast. \\
\indent Though the overall comparison between theoretical expectations and model results for dayside-nightside temperature differences in Figures \ref{fig:Acompare_high} and \ref{fig:Acompare_low} is useful, the observable dayside-nightside temperature differences are expected to be dominated by equatorial flow. Hence, we examine a comparison between model $A_{\mathrm{eq}}$ and the theoretical $A_{\mathrm{eq}}$ calculated from Equation (\ref{eqAtheory}). We do so for both the nonlinear case (see Figure \ref{fig:Acompare_eqhigh}) and linear case (Figure \ref{fig:Acompare_eqlow}). A similar general trend is seen as for the overall $A$, with the theory capturing well the change in $A_{\mathrm{eq}}$ with $\tau_{\mathrm{rad}}$ and $\tau_{\mathrm{drag}}$ and with the same mismatch in the weak-drag, strong-radiative cooling regime in the nonlinear limit. \\
\indent The quantitative comparison between theory and results is especially good in the $\Delta T_{\mathrm{eq,top}} = 0.001$ Kelvin case (Figure \ref {fig:Acompare_eqlow}). This increased compatibility is explained as in \cite{Perez-Becker:2013fv} by the lowered strength of horizontal advection as day-night forcing goes to zero. This causes the timescale comparison at the equator to be almost exactly vertical momentum advection against drag. Hence, in the linear limit the equatorial zonal-mean momentum equation becomes almost exactly pressure gradient force balancing either advection or drag. This enables the three-term analytic theory to predict dayside-nightside temperature differences most effectively, as we do not include the superrotating equatorial jet in our theory. \\
\indent Our theory also matches, to first order, the depth-dependence of dayside-nightside temperature differences over the range of radiative forcing and frictional drag strengths studied. Figure \ref{fig:apress} compares the pressure-dependent theoretical predictions for the dayside-nightside temperature difference with GCM results throughout the $\tau_{\mathrm{rad}}$, $\tau_{\mathrm{drag}}$ parameter space considered. This $\Delta T = T_{\mathrm{RMS,day}} - T_{\mathrm{RMS,night}}$ is a difference between the RMS dayside and RMS nightside temperatures, and hence is a representative dayside-nightside temperature difference, not the peak to trough difference. \\
\indent The same general trends are seen in both the theoretical predictions and numerical results, with $\Delta T$ being smallest at long $\tau_{\mathrm{rad}}$. As here the dominant comparison in the momentum equation is between pressure-gradient forces and either drag and Coriolis forces, the theoretical lines in a given column when $\tau_{\mathrm{drag}} \gtrsim 10^5 \hspace{2pt} \mathrm{sec}$ are equivalent. Referring back to \Eq{overallAtheory}, this means that drag does not have an effect on wave timescales unless it is stronger than the Coriolis force. That drag has a small effect on wave propagation timescales can also be seen in our numerical solutions by the similarity between the model curves for $\Delta T$ for $\tau_{\mathrm{drag}} \gtrsim 10^5 \hspace{2pt} \mathrm{sec}$. Hence, this is a re-iteration of the trend from Figure \ref{fig:nltempgrid} that $\tau_{\mathrm{rad}}$ is the main damping of wave adjustment processes in our model. Drag is a secondary effect that only becomes important at $\tau_{\mathrm{drag}} \lesssim 10^4$ sec. Crucially, the dayside-nightside temperature difference decreases with increasing pressure in both our theoretical predictions and numerical simulations. This is a key result which is due to the fact that $\tau_{\mathrm{rad}}$ increases with depth. Though this increase of the radiative timescale with depth is model-prescribed, it is expected from previous analytic \citep{showman_2002,Ginzburg:2015a} and numerical \citep{Iro:2005,Showman:2008} work which motivated this prescription.
\begin{figure*}[ht]
	\centering
	\includegraphics[width=.73\textwidth]{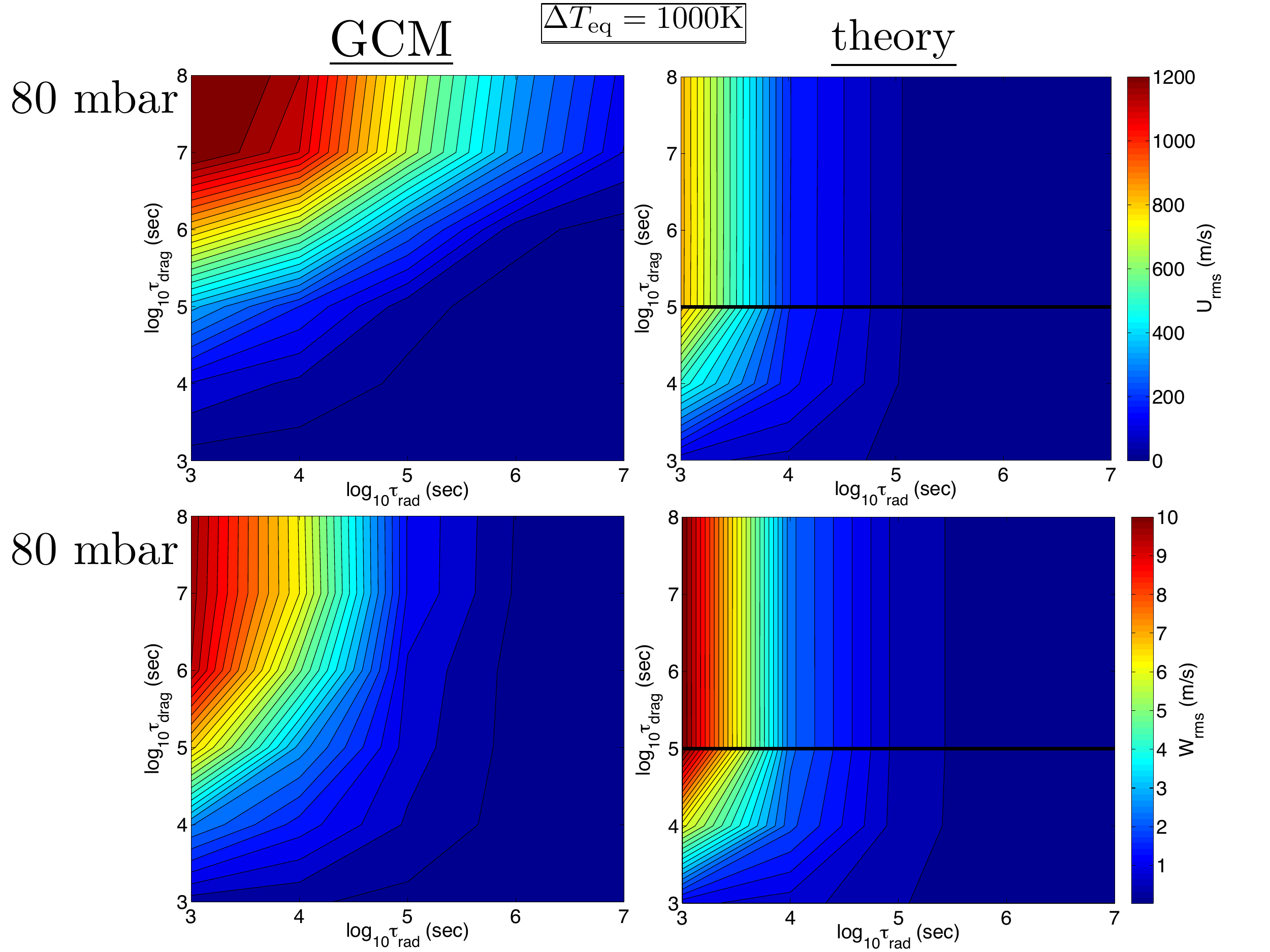}
	\caption{Root-mean-square velocities at 80 mbars, for model results (left) and theoretical prediction (right). The top row shows horizontal wind speeds, and the bottom row shows vertical wind speeds. The wind speeds are calculated using Equations (\ref{cont-simple}) and (\ref{eq:Wfinal}), with an overall three-way balance between Coriolis force, drag, and pressure gradient relevant to $\mathrm{Ro} < 1$.  The contours of wind speed are plotted as a function of $\tau_{\mathrm{rad,top}}$ (abscissa) and $\tau_{\mathrm{drag}}$ (ordinate), with the same contour scale for model and theory plots at a given pressure level. The black line divides the regions where Coriolis force dominates (above the line) and drag dominates (below the line). There is agreement within a factor of 2 for both horizontal and vertical wind speeds throughout almost all of parameter space.}
	\label{fig:windrms}
\end{figure*}
\subsection{Vertical and horizontal velocities}
\indent Using our predictive scaling for horizontal and vertical velocities, we compare GCM results with theoretical predictions for wind speeds in \Fig{fig:windrms}. Here, model and theory contours of velocities at 80 mbars pressure are plotted as a function of $\tau_{\mathrm{rad}}$ and $\tau_{\mathrm{drag}}$, as in the $A$ plots in Figures \ref{fig:Acompare_high} - \ref{fig:Acompare_eqlow}. The model results are RMS wind speeds, with $U_{\mathrm{rms}}$ defined in \Eq{eq:urms} and 
\begin{equation}
W_{\mathrm{rms}}(p) = \sqrt{\frac{\int w^2 dA}{A}} \mathrm{,}
\end{equation}
where $w$ is the vertical velocity at a given pressure level, calculated from our GCM results as $w = -\omega H/p$ \citep{Lewis:2010}. The theoretical predictions require only knowledge of the model input parameters (radiative equilibrium temperature profile, radiative and drag timescales). The same general trend of wind speeds increasing at shorter $\tau_{\mathrm{rad}}$ and longer $\tau_{\mathrm{drag}}$ is seen in both the model and theory plots. The theoretical wind speeds agree everywhere to within a factor of $\sim 2$. \\
\indent The largest discrepancy between theoretical and numerical horizontal wind speeds lies in the high $\tau_{\mathrm{rad}}$, high $\tau_{\mathrm{drag}}$ regime. This discrepancy is due to the fact that, in this corner of the parameter space, the GCM simulations exhibit longitudinal variations of temperature between day and night that are much smaller than the equator-to-pole temperature differences (see Figures \ref{fig:nltempgrid}-\ref{fig:lintempgrid}). In contrast, our theory only has one horizontal temperature difference--- we are implicitly assuming that the day-night temperature difference (as measured on latitude circles) is comparable to the equator-to-pole temperature difference. This assumption starts to become problematic in the high $\tau_{\mathrm{rad}}$, high $\tau_{\mathrm{drag}}$ portion of parameter space. In this region of parameter space, the equator-to-pole temperature difference is larger than the day-night temperature difference. This equator-to-pole temperature difference then drives strong horizontal winds, which are not predicted from our theory. This same discrepancy can be seen in the theory of \cite{Perez-Becker:2013fv}, see their Figure 6.  \\
\indent Given the simplicity of our theory, the agreement between predicted characteristic velocities and simulation RMS velocities is very good. This agreement is especially good for vertical velocities, because they are derived directly from the  thermodynamic weak temperature gradient balance. The horizontal velocities also agree within a factor of 2 over almost all of parameter space considered, except when drag and radiative forcing are both extremely weak. As a result, our velocity scaling can be used to predict characteristic velocities using dayside-nightside flux differences, which are attainable through phase curve observations. This is a powerful first determinant of the circulation of a given tidally locked exoplanet where the dayside-nightside temperature differences are larger than equator-to-pole temperature differences.
\begin{figure*}
	\centering
	\includegraphics[width=1\textwidth]{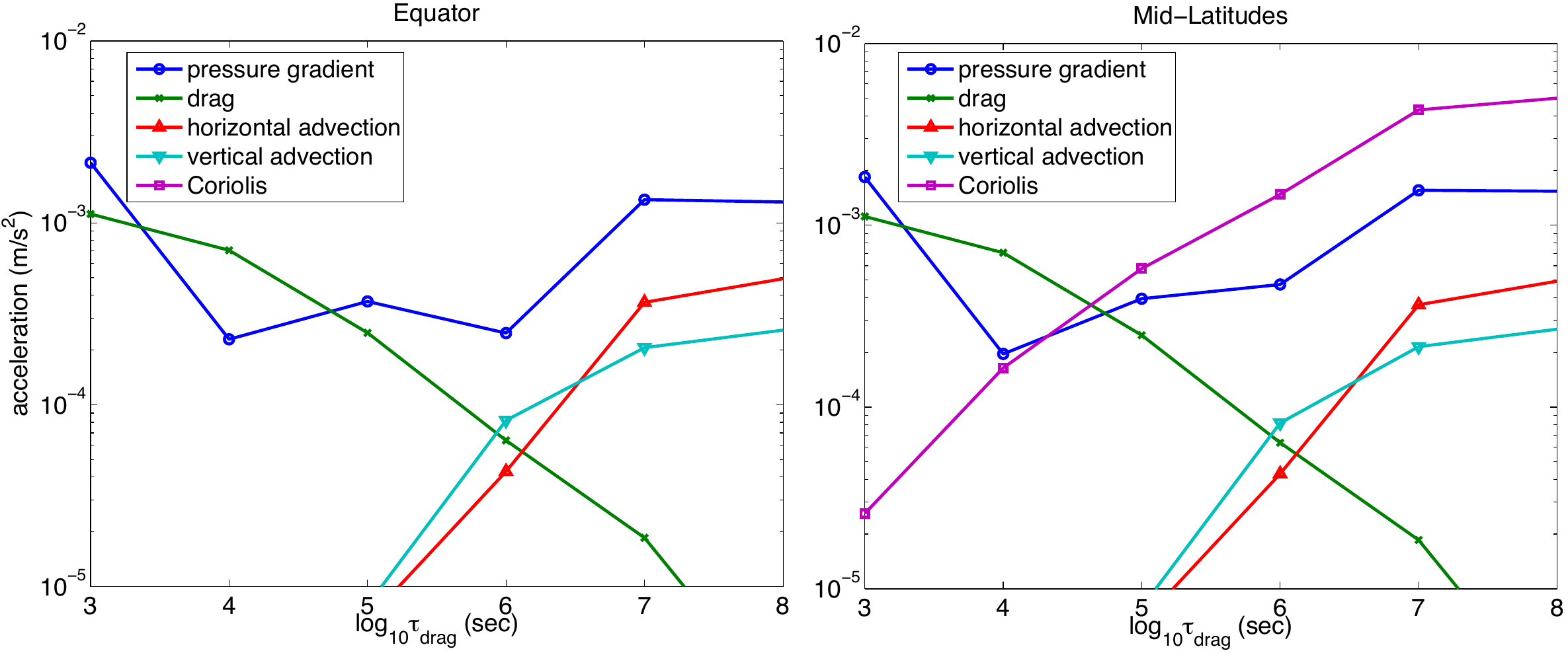}
	\caption{Comparison of absolute magnitudes of all terms in the approximate momentum equation, Equation (\ref{eq:horizmom}), as a function of $\tau_{\mathrm{drag}}$ for the $\Delta T_{\mathrm{eq,top}} = 1000$ Kelvin case with $\tau_{\mathrm{rad,top}} = 10^4$ sec at 80 mbars. The comparison is done both at the planetary equator ($0^{\circ}$ latitude, left) and mid-latitudes ($34^{\circ}$ latitude, right).}
	\label{mometermscomparison}
\end{figure*}
\begin{figure*}
	\centering
	\includegraphics[width=.75\textwidth]{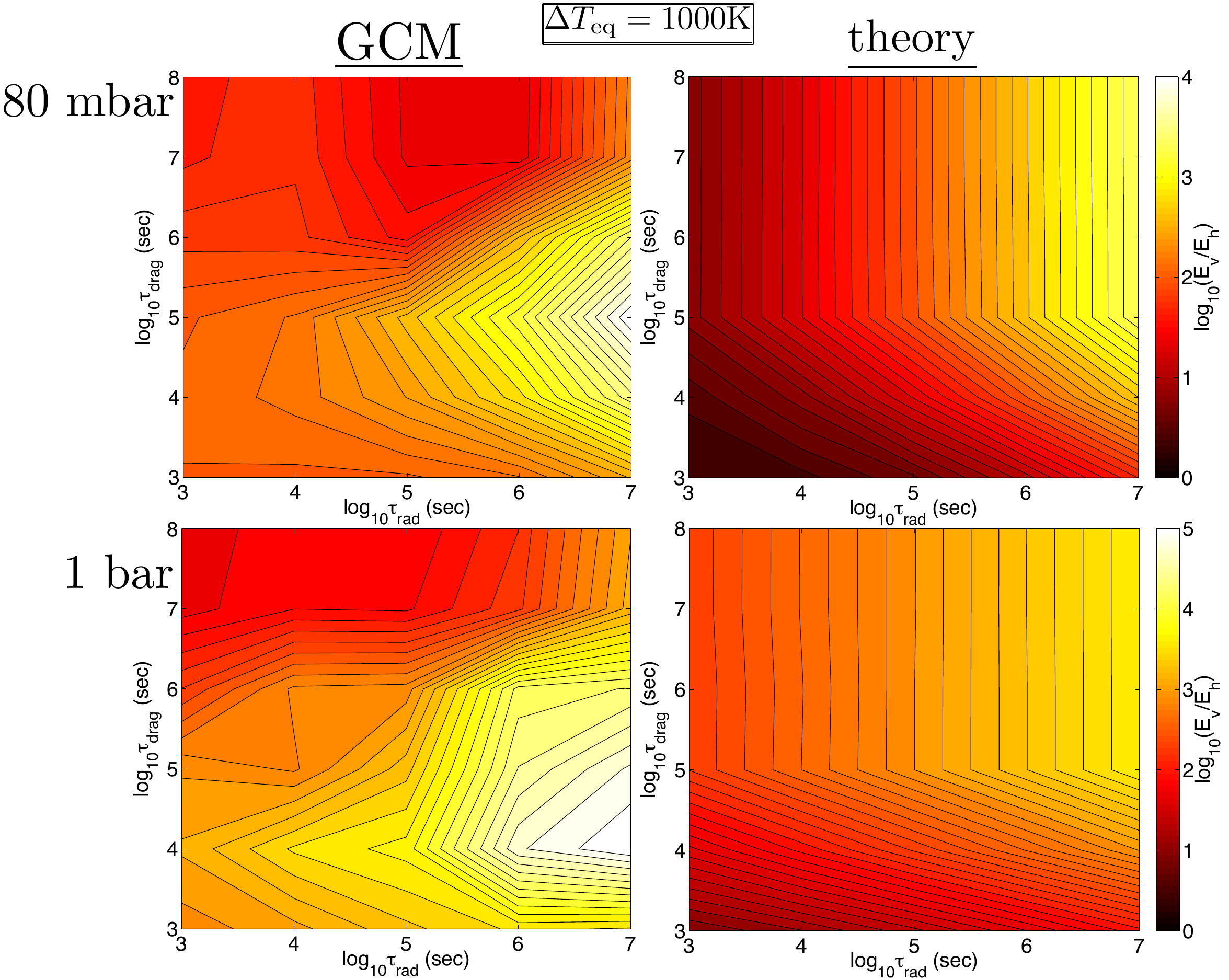}
	\caption{Ratio of vertical and horizontal entropy advection terms, plotted at pressures of 80 mbar (top) and 1 bar (bottom) for both GCM results (left) and theoretical prediction (right) as a function of $\tau_{\mathrm{rad,top}}$ (abscissa) and $\tau_{\mathrm{drag}}$ (ordinate). Model and theory plots for a given pressure level share a contour scale. The vertical entropy advection term is substantially larger than or comparable to the horizontal entropy advection term throughout all of the parameter space considered, in both the numerical results and theoretical predictions. This relative dominance of vertical entropy advection continues up to $\sim 10$ mbar pressure. Our choice of the weak temperature gradient regime for our analyses is therefore consistent with our numerical results.}
	\label{fig:etermscompare}
\end{figure*}

\subsection{Checks of regime determinations}
\subsubsection{Momentum equation}
\label{sec:dynamicalregimes}
\indent To confirm that the the dynamical regimes our model lies in agrees with theoretical predictions from \Sec{sec:momeqregime}, we examine the approximate momentum equation (Equation \ref{eq:horizmom}) at the equator and mid-latitudes. To calculate each term, we use the RMS values of each variable, setting $\mathcal{U} = U_{\mathrm{rms}}$, $\mathcal{W} = W_{\mathrm{rms}}$, and calculating $H$ using the RMS value of temperature at a given pressure level. Figure \ref{mometermscomparison} shows this comparison as a function of $\tau_{\mathrm{drag}}$. This shows the dominant balancing term with the pressure gradient force for the low and high drag regimes at both the equator and mid-latitudes. We choose $\phi = 34 ^{\circ}$ as a representative mid-latitude, as this is well northward of the region where the Coriolis forces and horizontal momentum advection terms are approximately equal, i.e. where $\mathrm{Ro} \approx 1$. This latitude is $\approx 0 - 25 ^{\circ}$ at $80 \ \mathrm{mbar}$ pressure, depending on the values of $\tau_{\mathrm{drag}}$ and $\tau_{\mathrm{rad}}$, as the values of these timescales strongly affect the wind speeds (see \Fig{fig:windrms}) and therefore the magnitude of the momentum advection term. Note that this is a comparison performed using our approximate momentum equation, and hence is a test both of our numerical results and theoretical approximations valid to the order-of-magnitude level. \\
\indent It is evident from Figure \ref{mometermscomparison} that drag balances the pressure gradient force for all $\tau_{\mathrm{drag}} \lesssim 10^5$ sec. At larger $\tau_{\mathrm{drag}}$, different forces balance pressure gradient at the equator and mid-latitudes. \\
\indent At the equator, horizontal and vertical momentum advection are the dominant balancing terms when $\tau_{\mathrm{drag}}$ is large. These approximate vertical and horizontal momentum advection terms are comparable over all $\tau_{\mathrm{rad}}$, $\tau_{\mathrm{drag}}$ parameter space, showing that our determination of the approximate continuity equation in \Eq{cont-simple}, which directly relates horizontal and vertical momentum advection, is appropriate.  \\
\indent At mid-latitudes, the Coriolis force is the dominant balancing term with pressure gradient when drag is weak, as it is orders of magnitude larger than the momentum advection terms. Hence, it was valid to consider two separate three-term force balances, at both the equator and mid-latitudes. At the equator, a three-way balance between advection, drag, and pressure gradient forces applies. At higher latitudes, the balance is between Coriolis, drag, and pressure gradient, as expected from our theory. 
\subsubsection{Energy equation}
\label{sec:wtg}
\indent Earth's tropics are in a weak temperature gradient regime, where vertical entropy advection dominates over horizontal entropy advection \citep{Sobel:2002}. In Section \ref{sec:thermo}, we determined using our analytical theory that the atmospheres of hot Jupiters lie in a similar regime. However, one might naively expect that horizontal entropy advection is more important in hot Jupiter atmospheres due to the large day to night forcing. \\
\indent To determine whether horizontal or vertical entropy advection is the controlling term in Equation (\ref{eq:energfirst}) over the $\tau_{\mathrm{rad}}$, $\tau_{\mathrm{drag}}$ parameter space studied, we can analyze our numerical solutions. Additionally, we examine the ratio of vertical and horizontal entropy advection terms from our analyses to show that the theory is self-consistent. To do so, we consider the approximate overall momentum balance between the Coriolis or drag forces and pressure gradient. This results in values of $A$ and hence $\Delta T$ given by \Eq{overallAtheory}. Figure \ref{fig:etermscompare} shows the approximate ratio of vertical and horizontal entropy advection terms, $E_v/E_h$: 
\begin{equation}
\label{eq:energyquot}
\frac{E_v}{E_h} \approx \frac{W_{\mathrm{rms}} N^2 H}{R} \left(\frac{U_{\mathrm{rms}} A \Delta T_{\mathrm{eq}}}{\mathcal{L}}\right)^{-1}\mathrm{.}
\end{equation}
This ratio is plotted as a function of $\tau_{\mathrm{rad}}$ and $\tau_{\mathrm{drag}}$ at two separate pressure levels (80 mbars and 1 bar) for both our numerical results and theoretical predictions. Note that we approximate $\Delta T \approx A \Delta T_{\mathrm{eq}}$, in order to calculate it consistently from our numerical solutions. \\
\indent Vertical entropy advection is comparable to or greater than horizontal entropy advection over the entire $\tau_{\mathrm{rad}},\tau_{\mathrm{drag}}$ parameter space, up to $\sim 10 \ \mathrm{mbars}$ pressure. The vertical entropy advection term is only comparable to horizontal entropy advection in our numerical results when $\tau_{\mathrm{drag}}$ is very large and $\tau_{\mathrm{rad}}$ short. Otherwise, the vertical entropy advection term is orders of magnitude larger than the horizontal entropy advection term. Additionally, our theory is self-consistent, as the predicted ratio of vertical to horizontal advection is always larger than unity. Hence, hot Jupiters, like the equatorial regions of Earth, lie largely in the weak temperature gradient regime---at least in a hemisphere-integrated sense (nevertheless, there will likely be local regions, particularly near the terminators, in which horizontal advection dominates).. This validates our use of the weak temperature gradient limit in our theory. This is the same limit found in \cite{Perez-Becker:2013fv} and similar to that used by \cite{Wordsworth:2014} to examine day-night differences on tidally-locked terrestrial exoplanets.
\section{Discussion}
\label{sec:discussion}
\indent As discussed in Section \ref{sec:wtg}, hot Jupiters lie in a weak temperature gradient regime, similar to the tropics of Earth \citep{Sobel:2001,Sobel:2002,Showman_2013_terrestrial_review}. In Earth's tropics, the propagation of gravity waves plays a key role in the adjustment of the thermal structure. Similarly, we interpret our theory to identify the mechanism that regulates dayside to nightside temperature differences on hot Jupiters as wave adjustment, analagous to the process that mediates temperature differences in Earth's tropics. This wave propagation can be damped, causing higher dayside-nightside temperature differences. This damping is largely due to radiative cooling, with a second order effect from potential atmospheric drag. Moreover, a simple theory (as presented in \Sec{sec:theory}) can explain the bulk of our model results for fractional dayside-nightside temperature differences over $\ge 4$ orders of magnitude variation in radiative and drag timescales. \\
\indent Additionally, these dayside-nightside temperature differences are a strong function of pressure. Namely, we expect that at depth the atmospheric circulation is more efficient at reducing day-night temperature differences. This depth-dependence is due in part to the vertical coupling between atmospheric levels but especially to the depth-dependence of radiative cooling, as $\tau_{\mathrm{rad}}$ increases with increasing pressure in our model, as justified by previous calculations of the radiative time constant \citep{Iro:2005,Showman:2008}. At pressures $\gtrsim 1$ bar, fractional dayside-nightside temperature differences are very small for the entire parameter regime studied, except in the case of extremely strong drag ($\tau_{\mathrm{drag}} \lesssim 10^4$ sec), or extremely short radiative timescales ($\tau_{\mathrm{rad,top}} \lesssim 10^4$ sec). \\
\indent The result that a competition between equatorial wave propagation and radiative cooling is the key balance mediating dayside-nightside temperature differences was previously found by \cite{Perez-Becker:2013fv}, and our results confirm theirs in 3D. More recently, \cite{Koll:2014} applied a similar understanding to terrestrial exoplanets. They found through numerical simulations of tidally-locked terrestrial planets that the ratio $\tau_{\mathrm{wave}}/\tau_{\mathrm{rad}}$ is a key determinant of dayside-nightside temperature differences. This is evident in our theory simply from how the fraction $\tau_{\mathrm{wave}}/\tau_{\mathrm{rad}}$ appears in every expression for $A(p)$ in \Eq{eq:tempfinal}. We expect that this comparison between wave and radiative timescales has consequences for the dayside-nightside temperature differences of a variety of tidally-locked gaseous exoplanets. Therefore, this comparison is applicable not only for the hot Jupiter regime but extends to terrestrial exoplanets.  \\
\indent However, one might note that unlike wave adjustment in response to cumulus convection in Earth's tropics \citep{Bretherton:1989}, which is a highly time-dependent process, we are here considering a steady forced/damped system, rather than examining the impact of freely propagating gravity waves. In the freely propagating case, these waves cause horizontal convergence/divergence, raising and lowering isentropes and causing lateral propagation with a (Kelvin) wave timescale $\tau_{\mathrm{wave}} \sim \mathcal{L}/NH$. Even through we do not consider freely propagating waves in our system, our solutions from \Sec{sec:theory} find the same natural wave timescale. This is unsurprising, as the strong heating on the dayside lowers isentropes relative to that on the nightside, leading to a system with isentropes that slope significantly from day to night. This sets up a horizontal pressure gradient force, and the dynamical response to this force causes flow that reduces the entropy difference between dayside and nightside at a given pressure level. This process is essentially the same physical mechanism as gravity wave adjustment, and likewise acts on a wave timescale \citep{Perez-Becker:2013fv}. \\
\indent The situation on a rotating planet has been examined by \cite{Perez-Becker:2013fv} (see their Section 5.3), which shows that there is a tight coupling between our numerical solutions and wave dynamics. The spatial structure of temperature in Figures \ref{fig:nltempgrid} and \ref{fig:lintempgrid} was shown to be due to standing, planetary-scale Rossby and Kelvin waves by \cite{Showman_Polvani_2011}. In the short $\tau_{\mathrm{rad}}$, long $\tau_{\mathrm{drag}}$ regime, one can identify the mid-latitude vorticial winds and temperature maximum as dynamically similar to a Rossby wave. Similarly, the equatorial divergence near the substellar point is analogous to a Kelvin wave. The standing solution in the linear limit represents behavior that is dynamically similar to the corresponding propagating modes, and is normally interpreted in terms of wave dynamics (e.g., \citealp{Matsuno:1966,Gill:1980}). Hence, we can re-iterate that the same mechanisms which cause propagation of these planetary-scale Kelvin and Rossby waves in the time-dependent case are the same mechanisms that regulate the steady-state dayside-nightside temperature differences in our hot Jupiter models. As such, the wave propagation timescale of a Kelvin wave is the fundamental wave timescale determining the transition from low to high day-to-night temperature differences. \\
\indent The importance of this Kelvin wave timescale for predicting the dayside-nightside temperature differences in hot Jupiter atmospheres was examined in \Sec{sec:transdT}. The transition from low-to-high dayside-nightside temperature differences necessarily depends on whether drag, Coriolis, or advection balances pressure-gradient forces in the momentum equation. However, we expect that over most latitudes Coriolis dominates over advection, and at temperatures $\gtrsim 1,500 \ \mathrm{K}$ where magnetic effects become important \citep{Perna_2010_1,Rogers:2014,Ginzburg:2015a} the day-night temperature differences will necessarily be large due to the very short $\tau_{\mathrm{rad}} \propto T^{-3}$ \citep{showman_2002}. Hence, referring back to \Eq{eq:transitionoverall}, a comparison between $\tau_{\mathrm{wave}}$ and $\sqrt{f^{-1} \tau_{\mathrm{rad}}(p)\Delta \mathrm{ln}p}$ determines whether day-night temperature differences will be low or high. If $\tau_{\mathrm{wave}}$ is shorter than $\sqrt{f^{-1} \tau_{\mathrm{rad}}(p)\Delta \mathrm{ln}p}$, the propagating tropical waves can extend longitudinally enough to reduce day-night temperature differences, leading to values of $A$ near zero. However, if $\sqrt{f^{-1} \tau_{\mathrm{rad}}(p)\Delta \mathrm{ln}p}$ is shorter than $\tau_{\mathrm{wave}}$, these equatorial Kelvin and Rossby waves are damped before they can zonally propagate, leading to high day-night temperature differences, with $A \sim 1$.  \\
\indent Our theory also enables estimation of horizontal and vertical velocities to within a factor of $\sim 2$ given dayside-nightside temperature differences. Notably, the velocity scaling for $\mathcal{U}$ in \Eq{eq:ucalc} allows estimation of horizontal velocities in the nominal hot Jupiter regime given only $\Delta T$ and $\Omega$, both of which can be calculated with phase curve information, assuming planet mass and radius are known. Additionally, our scaling for $\mathcal{W}$ in \Eq{eq:wcalc} enables estimation of vertical velocities given $\Delta T$, $\Omega$, and an estimate for Kelvin wave propagation timescales $\tau_{\mathrm{wave}}$ and radiative timescale $\tau_{\mathrm{rad}}$. These timescales can be computed with knowledge of the atmospheric composition and temperature. Hence, our theory enables first-order calculation of horizontal and vertical velocities directly from observable information, without the need for numerical simulations. This theory is a powerful tool for first insights on atmospheric circulation given only basic information about planet properties and dayside-nightside flux differences.
\section{Conclusions}
\label{sec:conclusions}
To summarize the main conclusions of our work from the discussion above:
\begin{enumerate}
\item Wave adjustment processes, analogous to those that mediate temperature differences in Earth's tropics, regulate dayside-to-nightside temperature differences in hot Jupiter atmospheres. The damping of wave adjustment processes in hot Jupiter atmospheres is mainly due to radiative cooling, with frictional drag playing a secondary role. Hot Jupiters have small dayside-nightside temperature differences at long $\tau_{\mathrm{rad}}$ and large dayside-nightside differences at short $\tau_{\mathrm{rad}}$, in accord with the observational trend of increasing dayside-nightside brightness temperature differences with increasing incident stellar flux.
\item When drag occurs on characteristic timescales faster than the rotation rate, dayside-nightside temperature differences are nominally large. Drag only has a second-order effect on dayside-nightside temperature differences in other cases, but does affect zonal wind speeds strongly, being a dominant mehanism to quell zonal winds.
\item We presented a simple analytic theory that explains reasonably well the day-night temperature differences and their dependence on radiative time constant, drag time constant, and pressure in three-dimensional models. Importantly, our analytic theory has zero free parameters, no tuning, and does not require any information from the GCM simulations to evaluate. This is the first fully predictive theory for the day-night temperature differences on hot Jupiters in three dimensions. Notably, a three-way force balance in conjunction with the weak temperature gradient limit can be used to describe dayside-nightside temperature differences at both the equator and higher latitudes. At the equator, the main forces balancing the day-to-night pressure gradient are momentum advection and drag, and at mid-latitudes drag and Coriolis forces balance the pressure gradient force. 
\item This analytic theory predicts that the transition between small and large day-night temperature differences is governed by a comparison between $\tau_{\mathrm{wave}}$ and $\sqrt{f^{-1} \tau_{\mathrm{rad}}(p)\Delta \mathrm{ln}p}$, with low day-night temperature differences if $\tau_{\mathrm{wave}} < \sqrt{f^{-1} \tau_{\mathrm{rad}}(p)\Delta \mathrm{ln}p}$ and high day-night temperature differences if $\tau_{\mathrm{wave}} > \sqrt{f^{-1} \tau_{\mathrm{rad}}(p)\Delta \mathrm{ln}p}$. If frictional drag has a characteristic timescale shorter than $1/f$, the transition is instead governed by a similar comparison between $\tau_{\mathrm{wave}}$ and $\sqrt{\tau_{\mathrm{drag}}(p)\tau_{\mathrm{rad}}(p)\Delta \mathrm{ln}p}$. The theory also covers the situation where both drag and Coriolis forces are weak compared to advective forces, a situation which should be relevant to planets that rotate especially slowly.
\item The same theory used to predict dayside-nightside temperature differences can also be inverted to calculate characteristic horizontal and vertical velocities from phase curve information. Horizontal velocities may be estimated given only the rotation rate and dayside-nightside flux differences on a given tidally locked planet. Similarly, vertical velocities may be calculated with the additional knowledge of atmospheric composition and temperature at a given pressure. This velocity scaling is a powerful tool to gain information on atmospheric circulation from data attainable through phase curves and may be extended to investigate properties of the circulation for the suite of tidally locked exoplanets without a boundary layer. 
\end{enumerate}
\acknowledgements
This research was supported by NASA Origins grant NNX12AI79G to APS. TDK acknowledges support from NASA headquarters under the NASA Earth and Space Science Fellowship Program Grant PLANET14F-0038. We thank the anonymous referee for useful comments. We also thank Josh Lothringer, Xianyu Tan, and Xi Zhang for helpful comments on the manuscript. Resources supporting this work were provided by the NASA High-End Computing (HEC) Program through the NASA Advanced Supercomputing (NAS) Division at Ames Research Center.
\appendix
\section{Observed Fractional Dayside-Nightside Brightness Temperature Differences}
\label{sec:append1}
\begin{table*}
\hspace{-0.0cm}
\begin{tabular}{| c | c | c | c | c | c |}
\hline
Planet & $T_{\mathrm{eq}}$ (K) & Dayside $T_b$ (K) & Dayside $T_b$ - Nightside $T_b$ (K) & $A_{\mathrm{obs}}$ & References \\
\hline
HD 189733b & $1202.8$ & $1250 \pm 13$ & $313 \pm 23$ & $0.250 \pm 0.018$ & \cite{Knutson_2007,Knutson:2009,Knutson:2012} \\
\hline
WASP-43b & $1378.0$ & $1725 \pm 6$ & N/A &  $\ge 0.54$ & \cite{Stevenson:2014} \\
\hline
HD 209458b & $1449.5$ & $1499 \pm 15$ & $527 \pm 46$ & $0.352 \pm 0.031$ & \cite{Crossfield:2012,Zellem:2014} \\
\hline
HD 149026b & $1677.8$ & $1440 \pm 150$ & $480 \pm 140$ & $0.333 \pm 0.103$ & \cite{Knutson:2009a} \\
\hline
WASP-14b & $1869.2$ & $2302 \pm 38$ & $1081 \pm 82.3$ & $0.470 \pm 0.037$ & \cite{Wong:2015} \\
\hline
WASP-19b & $2070.2$ & $2357 \pm 64$ & $1227 \pm 248$ & $0.521 \pm 0.106$ & \cite{Wong:2015a} \\
\hline
HAT-P-7b & $2230.0$ & $2682 \pm 49$ & $972 \pm 187$ & $0.362 \pm 0.070$ & \cite{Wong:2015a} \\
\hline
WASP-18b & $2403.3$ & $3153 \pm 51$ & N/A & $\ge 0.40$ & \cite{Nymeyer:2011,Maxted:2013} \\
\hline
WASP-12b & $2589.7$ & $2928 \pm 97$ & $1945 \pm 223$ & $0.664 \pm 0.079$ & \cite{Cowan:2012} \\
\hline
\end{tabular}
\label{table:aobs}
\caption{Compilation of data used to make Figure \ref{fig:Aobs}, along with appropriate references.}
\end{table*}
Table 1 displays the data utilized to calculate the $A_{\mathrm{obs}}-T_{\mathrm{eq}}$ points shown in Figure \ref{fig:Aobs}, along with appropriate references. To compile this data, we collected either the error-weighted or photometric dayside brightness temperature and dayside-nightside brightness temperature differences. If no error-weighted values were given, we calculated each brightness temperature as an arithmetic mean of that provided at each wavelength. Specifically, we computed arithmetic means for HD 189733b, WASP-43b, and WASP-18b, with the other transiting planets either having an error-weighted value provided or only one photometric wavelength with available data. For both WASP-19b and HAT-P-7b, we utilized the $4.5 \mu\mathrm{m}$ day-night brightness temperature difference, as only upper limits on the nightside brightness temperatures at $3.6 \mu\mathrm{m}$ were available. Note that though HAT-P-7b seems to have a day-night brightness temperature difference somewhat below that expected, the lower limit on $A_{\mathrm{obs}}$ at $3.6 \mu\mathrm{m}$ is $0.483$ \citep{Wong:2015a}, $33.4\%$ larger than the value of $A_{\mathrm{obs}}$ at $4.5 \mu\mathrm{m}$.   \\
\indent To compute a lower limit on $A_{\mathrm{obs}}$ for WASP-43b, we utilized the $1\sigma$ upper limit on the nightside bolometric flux from \cite{Stevenson:2014}. We used a similar method for WASP-18b, averaging the lower limits on $A_{\mathrm{obs}}$ from the $3.6\mu$m and $4.5 \mu$m nightside $1\sigma$ flux upper limits from \cite{Maxted:2013}. This enabled us to calculate the lower limit of $A_{\mathrm{obs}} \equiv A_{\mathrm{obs,ll}}$ as a function of dayside and nightside fluxes as
\begin{equation}
\label{eq:aobsflux}
A_{\mathrm{obs,ll}} = \frac{F_{\mathrm{day}}^{1/4} - F_{\mathrm{night,max}}^{1/4}}{F_{\mathrm{day}}^{1/4}} \mathrm{.}
\vspace{0.01cm}
\end{equation}
In \Eq{eq:aobsflux} above, $F_{\mathrm{day}}$ is the dayside bolometric flux and $F_{\mathrm{night,max}}$ is the upper limit on the nightside bolometric flux. 

\if\bibinc n
\bibliography{References}

\begin{thebibliography}{73}
\expandafter\ifx\csname natexlab\endcsname\relax\def\natexlab#1{#1}\fi

\bibitem[{Adcroft {et~al.}(2004)Adcroft, Hill, Campin, Marshall, \&
  Heimbach}]{Adcroft:2004}
Adcroft, A., Hill, C., Campin, J., Marshall, J., \& Heimbach, P. 2004, Monthly
  Weather Review, 132, 2845

\bibitem[{Andrews {et~al.}(1987)Andrews, Holton, \& Leovy}]{Andrews:1987}
Andrews, D., Holton, J., \& Leovy, C. 1987, International Geophysics Series,
  Vol.~40, Middle Atmosphere Dynamics, ed. R.~Dmowska \& J.~Holton (Academic
  Press)

\bibitem[{Batygin {et~al.}(2013)Batygin, Stanley, \& Stevenson}]{batygin_2013}
Batygin, K., Stanley, S., \& Stevenson, D. 2013, The Astrophysical Journal,
  776, 53

\bibitem[{Batygin \& Stevenson(2010)}]{Batygin_2010}
Batygin, K. \& Stevenson, D. 2010, The Astrophysical Journal Letters, 714, L238

\bibitem[{Borucki {et~al.}(2009)Borucki, Koch, Jenkins, Sasselov, Gilliland,
  Batalha, Latham, Caldwell, Basri, Brown, Christensen-Dalsgaard, Cochran,
  Devore, Dunham, Dupree, Gautier, Geary, Gould, Howell, Kjeldsen, Lissauer,
  Marcy, Meibom, Morrison, \& Tarter}]{Borucki:2009}
Borucki, W., Koch, D., Jenkins, J., Sasselov, D., Gilliland, R., Batalha, N.,
  Latham, D., Caldwell, D., Basri, G., Brown, T., Christensen-Dalsgaard, J.,
  Cochran, W., Devore, E., Dunham, D., Dupree, A., Gautier, T., Geary, J.,
  Gould, A., Howell, S., Kjeldsen, H., Lissauer, J., Marcy, G., Meibom, S.,
  Morrison, D., \& Tarter, J. 2009, Science, 325, 709

\bibitem[{Bretherton \& Smolarkiewicz(1989)}]{Bretherton:1989}
Bretherton, C. \& Smolarkiewicz, P. 1989, Journal of the Atmospheric Sciences,
  46, 740

\bibitem[{Charbonneau {et~al.}(2000)Charbonneau, Brown, Latham, \&
  Mayor}]{Charbonneau_2000}
Charbonneau, D., Brown, T., Latham, D., \& Mayor, M. 2000, The Astrophysical
  Journal, 529, L45

\bibitem[{Cooper \& Showman(2005)}]{Cooper:2005}
Cooper, C. \& Showman, A. 2005, The Astrophysical Journal Letters, 629, L45

\bibitem[{Cowan \& Agol(2011)}]{Cowan_2011}
Cowan, N. \& Agol, E. 2011, The Astrophysical Journal, 759, 54

\bibitem[{Cowan {et~al.}(2007)Cowan, Agol, \& Charbonneau}]{Cowan:2007}
Cowan, N., Agol, E., \& Charbonneau, D. 2007, Monthly Notices of the Royal
  Astronomical Society, 379, 641

\bibitem[{Cowan {et~al.}(2012)Cowan, Machalek, Croll, Shekhtman, Burrows,
  Deming, Greene, \& Hora}]{Cowan:2012}
Cowan, N., Machalek, P., Croll, B., Shekhtman, L., Burrows, A., Deming, D.,
  Greene, T., \& Hora, J. 2012, The Astrophysical Journal, 747, 82

\bibitem[{Crossfield {et~al.}(2012)Crossfield, Knutson, Fortney, Showman,
  Cowan, \& Deming}]{Crossfield:2012}
Crossfield, I., Knutson, H., Fortney, J., Showman, A., Cowan, N., \& Deming, D.
  2012, The Astrophysical Journal, 752, 81

\bibitem[{Demory {et~al.}(2013)Demory, Wit, Lewis, Fortney, Zsom, Seager,
  Knutson, Heng, Madhusudhan, Gillon, Barclay, Desert, Parmentier, \&
  Cowan}]{Demory_2013}
Demory, B., Wit, J.~D., Lewis, N., Fortney, J., Zsom, A., Seager, S., Knutson,
  H., Heng, K., Madhusudhan, N., Gillon, M., Barclay, T., Desert, J.,
  Parmentier, V., \& Cowan, N. 2013, The Astrophysical Journal Letters, 776,
  L25

\bibitem[{Dobbs-Dixon \& Agol(2013)}]{Dobbs-Dixon:2013}
Dobbs-Dixon, I. \& Agol, E. 2013, Monthly Notices of the Royal Astronomical
  Society, 435, 3159

\bibitem[{{Fortney} {et~al.}(2008){Fortney}, {Lodders}, {Marley}, \&
  {Freedman}}]{fortney-etal-2008}
{Fortney}, J.~J., {Lodders}, K., {Marley}, M.~S., \& {Freedman}, R.~S. 2008,
  \apj, 678, 1419

\bibitem[{Gill(1980)}]{Gill:1980}
Gill, A. 1980, Quarterly Journal of the Royal Meteorological Society, 106, 447

\bibitem[{Ginzburg \& Sari(2015)}]{Ginzburg:2015a}
Ginzburg, S. \& Sari, R. 2015, arXiv 1511.00135

\bibitem[{{Held}(2005)}]{held-2005}
{Held}, I. 2005, Bull. Amer. Meteorological Soc., 86, 1609

\bibitem[{Held \& Suarez(1994)}]{Held:1994}
Held, I. \& Suarez, M. 1994, Bulletin of the American Meteorological Society,
  75, 1825

\bibitem[{Heng {et~al.}(2011{\natexlab{a}})Heng, Frierson, \&
  Phillips}]{Heng:2011a}
Heng, K., Frierson, D., \& Phillips, P. 2011{\natexlab{a}}, Monthly Notices of
  the Royal Astronomical Society, 418, 2669

\bibitem[{Heng {et~al.}(2011{\natexlab{b}})Heng, Menou, \&
  Phillips}]{Heng:2011}
Heng, K., Menou, K., \& Phillips, P. 2011{\natexlab{b}}, Monthly Notices of the
  Royal Astronomical Society, 413, 2380

\bibitem[{Henry {et~al.}(2000)Henry, Marcy, Butler, \& Vogt}]{Henry:2000}
Henry, G., Marcy, G., Butler, R., \& Vogt, S. 2000, The Astrophysical Journal,
  529, L41

\bibitem[{Holton \& Hakim(2013)}]{Holton:2013}
Holton, J. \& Hakim, G. 2013, An introduction to dynamic meteorology, 5th edn.
  (Academic Press)

\bibitem[{Iro {et~al.}(2005)Iro, B\'{e}zard, \& Guillot}]{Iro:2005}
Iro, N., B\'{e}zard, B., \& Guillot, T. 2005, Astronomy and Astrophysics, 436,
  719

\bibitem[{Kataria {et~al.}(2015)Kataria, Showman, Fortney, Stevenson, Line,
  Kriedberg, Bean, \& Desert}]{Kataria:2014}
Kataria, T., Showman, A., Fortney, J., Stevenson, K., Line, M., Kriedberg, L.,
  Bean, J., \& Desert, J. 2015, The Astrophysical Journal, 801, 86

\bibitem[{Kataria {et~al.}(2013)Kataria, Showman, Lewis, Fortney, Marley, \&
  Freedman}]{kataria_2013}
Kataria, T., Showman, A., Lewis, N., Fortney, J., Marley, M., \& Freedman, R.
  2013, The Astrophysical Journal, 767, 76

\bibitem[{Knutson {et~al.}(2007)Knutson, Charbonneau, Allen, Fortney, Agol,
  Cowan, Showman, Cooper, \& Megeath}]{Knutson_2007}
Knutson, H., Charbonneau, D., Allen, L., Fortney, J., Agol, E., Cowan, N.,
  Showman, A., Cooper, C., \& Megeath, S. 2007, Nature, 447, 183

\bibitem[{Knutson {et~al.}(2009{\natexlab{a}})Knutson, Charbonneau, Cowan,
  Fortney, Showman, Agol, \& Henry}]{Knutson:2009a}
Knutson, H., Charbonneau, D., Cowan, N., Fortney, J., Showman, A., Agol, E., \&
  Henry, G. 2009{\natexlab{a}}, The Astrophysical Journal, 703, 769

\bibitem[{Knutson {et~al.}(2009{\natexlab{b}})Knutson, Charbonneau, Cowan,
  Fortney, Showman, Agol, Henry, Everett, \& Allen}]{Knutson:2009}
Knutson, H., Charbonneau, D., Cowan, N., Fortney, J., Showman, A., Agol, E.,
  Henry, G., Everett, M., \& Allen, L. 2009{\natexlab{b}}, The Astrophysical
  Journal, 690, 822

\bibitem[{Knutson {et~al.}(2012)Knutson, Lewis, Fortney, Burrows, Showman,
  Cowan, Agol, Aigrain, Charbonneau, Deming, Desert, Henry, Langton, \&
  Laughlin}]{Knutson:2012}
Knutson, H., Lewis, N., Fortney, J., Burrows, A., Showman, A., Cowan, N., Agol,
  E., Aigrain, S., Charbonneau, D., Deming, D., Desert, J., Henry, G., Langton,
  J., \& Laughlin, G. 2012, The Astrophysical Journal, 754, 22

\bibitem[{Koll \& Abbot(2015)}]{Koll:2014}
Koll, D. \& Abbot, D. 2015, The Astrophysical Journal, 802, 21

\bibitem[{Lewis {et~al.}(2010)Lewis, Showman, Fortney, Marley, Freedman, \&
  Lodders}]{Lewis:2010}
Lewis, N., Showman, A., Fortney, J., Marley, M., Freedman, R., \& Lodders, K.
  2010, The Astrophysical Journal, 720, 344

\bibitem[{Li \& Goodman(2010)}]{Li:2010}
Li, J. \& Goodman, J. 2010, The Astrophysical Journal, 725, 1146

\bibitem[{Liu \& Showman(2013)}]{Liu:2013}
Liu, B. \& Showman, A. 2013, The Astrophysical Journal, 770, 42

\bibitem[{Matsuno(1966)}]{Matsuno:1966}
Matsuno, T. 1966, Journal of the Meteorological Society of Japan, 44, 25

\bibitem[{Maxted {et~al.}(2013)Maxted, Anderson, Doyle, Gillon, Harrington,
  Iro, Jehin, Lafreniere, Smalley, \& Southworth}]{Maxted:2013}
Maxted, P., Anderson, D., Doyle, A., Gillon, M., Harrington, J., Iro, N.,
  Jehin, E., Lafreniere, D., Smalley, B., \& Southworth, J. 2013, Monthly
  Notices of the Royal Astronomical Society, 428, 2645

\bibitem[{Mayne {et~al.}(2014)Mayne, Baraffe, Acreman, Smith, Browning,
  Amundsen, Wood, Thuburn, \& Jackson}]{Mayne:2014}
Mayne, N., Baraffe, I., Acreman, D., Smith, C., Browning, M., Amundsen, D.,
  Wood, N., Thuburn, J., \& Jackson, D. 2014, Astronomy and Astrophysics, 561,
  A1

\bibitem[{Menou(2012)}]{Menou:2012fu}
Menou, K. 2012, The Astrophysical Journal, 745, 138

\bibitem[{Menou \& Rauscher(2009)}]{Menou:2009}
Menou, K. \& Rauscher, E. 2009, The Astrophysical Journal, 700, 887

\bibitem[{Nymeyer {et~al.}(2011)Nymeyer, Harrington, Hardy, Stevenson, Campo,
  Madhusudhan, Collier-Cameron, Loredo, Blecic, Bowman, Britt, Cubillos,
  Hellier, Gillon, Maxted, Hebb, Wheatley, Pollaco, \& Anderson}]{Nymeyer:2011}
Nymeyer, S., Harrington, J., Hardy, R., Stevenson, K., Campo, C., Madhusudhan,
  N., Collier-Cameron, A., Loredo, T., Blecic, J., Bowman, W., Britt, C.,
  Cubillos, P., Hellier, C., Gillon, M., Maxted, P., Hebb, L., Wheatley, P.,
  Pollaco, D., \& Anderson, D. 2011, The Astrophysical Journal, 742, 35

\bibitem[{Parmentier {et~al.}(2013)Parmentier, Showman, \&
  Lian}]{parmentier_2013}
Parmentier, V., Showman, A., \& Lian, Y. 2013, Astronomy and Astrophysics, 558,
  A91

\bibitem[{Perez-Becker \& Showman(2013)}]{Perez-Becker:2013fv}
Perez-Becker, D. \& Showman, A. 2013, The Astrophysical Journal, 776, 134

\bibitem[{Perna {et~al.}(2012)Perna, Heng, \& Pont}]{perna_2012}
Perna, R., Heng, K., \& Pont, F. 2012, The Astrophysical Journal, 751, 59

\bibitem[{Perna {et~al.}(2010)Perna, Menou, \& Rauscher}]{Perna_2010_1}
Perna, R., Menou, K., \& Rauscher, E. 2010, The Astrophysical Journal, 719,
  1421

\bibitem[{Polvani \& Sobel(2001)}]{Polvani:2001}
Polvani, L. \& Sobel, A. 2001, Journal of the Atmospheric Sciences, 59, 1744

\bibitem[{Rauscher \& Menou(2010)}]{Rauscher:2010}
Rauscher, E. \& Menou, K. 2010, The Astrophysical Journal, 714, 1334

\bibitem[{Rauscher \& Menou(2012{\natexlab{a}})}]{Rauscher_2012}
---. 2012{\natexlab{a}}, The Astrophysical Journal, 750, 96

\bibitem[{Rauscher \& Menou(2012{\natexlab{b}})}]{Rauscher:2012}
---. 2012{\natexlab{b}}, The Astrophysical Journal, 745, 78

\bibitem[{Rauscher \& Menou(2013)}]{Rauscher_2013}
---. 2013, The Astrophysical Journal, 764, 103

\bibitem[{Rogers \& Komacek(2014)}]{Rogers:2014}
Rogers, T. \& Komacek, T. 2014, The Astrophysical Journal, 794, 132

\bibitem[{Rogers \& Showman(2014)}]{Rogers:2020}
Rogers, T. \& Showman, A. 2014, The Astrophysical Journal Letters, 782, L4

\bibitem[{Schwartz \& Cowan(2015)}]{Schwartz:2015}
Schwartz, J. \& Cowan, N. 2015, Monthly Notices of the Royal Astronomical
  Society, 449, 4192

\bibitem[{Showman {et~al.}(2010)Showman, Cho, \& Menou}]{Showman_2009}
Showman, A., Cho, J., \& Menou, K. Exoplanets, ed. S.~Seager (Tucson, AZ:
  University of Arizona Press)

\bibitem[{Showman {et~al.}(2008{\natexlab{a}})Showman, Cooper, Fortney, \&
  Marley}]{Showman:2008}
Showman, A., Cooper, C., Fortney, J., \& Marley, M. 2008{\natexlab{a}}, The
  Astrophysical Journal, 682, 559

\bibitem[{Showman {et~al.}(2013{\natexlab{a}})Showman, Fortney, Lewis, \&
  Shabram}]{showman_2013_doppler}
Showman, A., Fortney, J., Lewis, N., \& Shabram, M. 2013{\natexlab{a}}, The
  Astrophysical Journal, 762, 24

\bibitem[{Showman {et~al.}(2009)Showman, Fortney, Lian, Marley, Freedman,
  Knutson, \& Charbonneau}]{Showmanetal_2009}
Showman, A., Fortney, J., Lian, Y., Marley, M., Freedman, R., Knutson, H., \&
  Charbonneau, D. 2009, The Astrophysical Journal, 699, 564

\bibitem[{Showman \& Guillot(2002)}]{showman_2002}
Showman, A. \& Guillot, T. 2002, Astronomy and Astrophysics, 385, 166

\bibitem[{Showman {et~al.}(2015)Showman, Lewis, \& Fortney}]{Showman:2014}
Showman, A., Lewis, N., \& Fortney, J. 2015, The Astrophysical Journal, 801, 95

\bibitem[{Showman {et~al.}(2008{\natexlab{b}})Showman, Menou, \&
  Cho}]{showman_2008}
Showman, A., Menou, K., \& Cho, J. 2008{\natexlab{b}}, Extreme Solar Systems
  ASP Conference Series, 398, 419

\bibitem[{Showman \& Polvani(2010)}]{Showman:2010}
Showman, A. \& Polvani, L. 2010, Geophysical Research Letters, 37, L18811

\bibitem[{Showman \& Polvani(2011)}]{Showman_Polvani_2011}
---. 2011, The Astrophysical Journal, 738, 71

\bibitem[{Showman {et~al.}(2013{\natexlab{b}})Showman, Wordsworth, Merlis, \&
  Kaspi}]{Showman_2013_terrestrial_review}
Showman, A., Wordsworth, R., Merlis, T., \& Kaspi, Y. Comparative Climatology
  of Terrestrial Planets, ed. S.~Mackwell, A.~Simon-Miller, J.~Harder, \&
  M.~Bullock (Tucson, AZ: University of Arizona Press)

\bibitem[{Sobel(2002)}]{Sobel:2002}
Sobel, A. 2002, Chaos, 12, 451

\bibitem[{Sobel {et~al.}(2001)Sobel, Nilsson, \& Polvani}]{Sobel:2001}
Sobel, A., Nilsson, J., \& Polvani, L. 2001, Journal of the Atmospheric
  Sciences, 58, 3650

\bibitem[{Stevenson {et~al.}(2014)Stevenson, Desert, Line, Bean, Fortney,
  Showman, Kataria, Kriedberg, McCullough, Henry, Charbonneau, Burrows, Seager,
  Madhusudhan, Williamson, \& Homeier}]{Stevenson:2014}
Stevenson, K., Desert, J., Line, M., Bean, J., Fortney, J., Showman, A.,
  Kataria, T., Kriedberg, L., McCullough, P., Henry, G., Charbonneau, D.,
  Burrows, A., Seager, S., Madhusudhan, N., Williamson, M., \& Homeier, D.
  2014, Science, 346, 838

\bibitem[{Thrastarson \& Cho(2010)}]{Thrastarson:2010}
Thrastarson, H. \& Cho, J. 2010, The Astrophysical Journal, 716, 144

\bibitem[{Tsai {et~al.}(2014)Tsai, Dobbs-Dixon, \& Gu}]{Tsai:2014}
Tsai, S., Dobbs-Dixon, I., \& Gu, P. 2014, The Astrophysical Journal, 793, 141

\bibitem[{Wallace \& Hobbs(2004)}]{Wallace:2004}
Wallace, J. \& Hobbs, P. 2004, Atmospheric Science: An Introductory Survey, 2nd
  edn. (San Diego: Academic Press)

\bibitem[{Wong {et~al.}(2015{\natexlab{a}})Wong, Knutson, Lewis, Burrows, Agol,
  Cowan, Deming, Desert, Fortney, Fulton, Howard, Kataria, Langton, Laughlin,
  Schwartz, Showman, \& Todorov}]{Wong:2015}
Wong, I., Knutson, H., Lewis, N., Burrows, A., Agol, E., Cowan, N., Deming, D.,
  Desert, J., Fortney, J., Fulton, B., Howard, A., Kataria, T., Langton, J.,
  Laughlin, G., Schwartz, J., Showman, A., \& Todorov, K. 2015{\natexlab{a}},
  The Astrophysical Journal, 811, 122

\bibitem[{Wong {et~al.}(2015{\natexlab{b}})Wong, Knutson, Lewis, Kataria,
  Burrows, Fortney, Schwartz, Shporer, Agol, Cowan, Deming, Desert, Fulton,
  Howard, Langton, Laughlin, Showman, \& Todorov}]{Wong:2015a}
Wong, I., Knutson, H., Lewis, N., Kataria, T., Burrows, A., Fortney, J.,
  Schwartz, J., Shporer, A., Agol, E., Cowan, N., Deming, D., Desert, J.,
  Fulton, B., Howard, A., Langton, J., Laughlin, G., Showman, A., \& Todorov,
  K. 2015{\natexlab{b}}, arXiv 1512.09342

\bibitem[{Wordsworth(2015)}]{Wordsworth:2014}
Wordsworth, R. 2015, The Astrophysical Journal, 806, 180

\bibitem[{Youdin \& Mitchell(2010)}]{Youdin_2010}
Youdin, A. \& Mitchell, J. 2010, The Astrophysical Journal, 721, 1113

\bibitem[{Zellem {et~al.}(2014)Zellem, Lewis, Knutson, Griffith, Showman,
  Fortney, Cowan, Agol, Burrows, Charbonneau, Deming, Laughlin, \&
  Langton}]{Zellem:2014}
Zellem, R., Lewis, N., Knutson, H., Griffith, C., Showman, A., Fortney, J.,
  Cowan, N., Agol, E., Burrows, A., Charbonneau, D., Deming, D., Laughlin, G.,
  \& Langton, J. 2014, The Astrophysical Journal, 790, 53

\end{thebibliography}


\begin{thebibliography}
\end{thebibliography}
\fi

\if\bibinc y

\fi

\end{document}